# Unleashing the power of computational insights in revealing the complexity of biological systems in the new era of spatial multi-omics

Zhiwei Fan[1], Tiangang Wang[2], Kexin Huang[2,3,*], Binwu Ying[4,5,*], Xiaobo Zhou[6,7,*]

[1]West China Biomedical Big Data Center, West China Hospital; Med-X Center for Informatics, Sichuan University, Chengdu, 610044, China.

[2]School of Life Science and Technology, Xidian University, Xi'an, Shaanxi 710071, P.R. China

[3]Center for Computational Systems Medicine, School of Biomedical Informatics, The University of Texas Health Science Center at Houston, Houston, TX 77030, USA

[4]Department of Laboratory Medicine, West China Hospital, Sichuan University, Chengdu 610041, China

[5]Clinical Laboratory Medicine Research Center, West China Hospital, Sichuan University, Chengdu 610041, China Sichuan Clinical Research Center for Laboratory Medicine, Chengdu 610041, China.

[6]McGovern Medical School, The University of Texas Health Science Center at Houston, Houston, TX 77030, USA

[7]School of Dentistry, The University of Texas Health Science Center at Houston, Houston, TX 77030, USA

## Abstract

Recent advances in spatial omics technologies have revolutionized our ability to study biological systems with unprecedented resolution. By preserving the spatial context of molecular measurements, these methods enable comprehensive mapping of cellular heterogeneity, tissue architecture, and dynamic biological processes in developmental biology, neuroscience, oncology, and evolutionary studies. This review highlights a systematic overview of the continuous advancements in both technology and computational algorithms that are paving the way for a deeper, more systematic comprehension of the structure and mechanisms of mammalian tissues and organs by using spatial multi-omics. Our viewpoint demonstrates how advanced machine learning algorithms and multi-omics integrative modeling can decode complex biological processes, including the spatial organization and topological relationships of cells during organ development, as well as key molecular signatures and regulatory networks underlying tumorigenesis and metastasis. Finally, we outline future directions for technological innovation and modeling insights of spatial omics in precision medicine.

## 1. Introduction

Over the past two decades, high-throughput sequencing technologies have revolutionized our understanding of biological systems by enabling comprehensive molecular profiling of tissues and cells. The initial breakthroughs in transcriptomics, notably bulk RNA-seq, facilitated the quantification of global gene expression patterns across diverse conditions, tissues, and disease states[1, 2]. While bulk RNA-seq provided a powerful foundation for characterizing the transcriptome, it lacked the resolution to disentangle cellular heterogeneity and was inherently limited in capturing the spatial organization of gene expression within complex tissues.

The emergence of single-cell RNA sequencing revolutionized this paradigm. Single-cell transcriptomics, awarded Method of the Year status in 2013[3], enabled the dissection of transcriptional programs at cellular resolution. It has revealed a remarkable diversity of cell types and states, transformed developmental biology, and provided critical insights into the cellular ecosystems of organs, tumors, and immune responses[4-7]. However, transcriptomes alone provide a limited view of cellular biology, as they reflect one regulatory layer and are influenced by transcriptional noise, RNA turnover, and post-transcriptional processes. This realization led to the development of single-cell multi-omics, which integrates multiple molecular layers—such as chromatin accessibility, DNA methylation, surface protein abundance, and transcriptomes—within individual cells[8]. Highlighted as Method of the Year in 2019[9], these approaches offered a more holistic depiction of gene regulatory logic by linking genotype to phenotype across multiple biological dimensions[10-12].

However, these methods still required tissue dissociation, thereby removing cells from their native spatial context. To address this limitation, attention shifted toward spatial transcriptomics, which preserves the in situ localization of gene expression within intact tissue architecture[13-15]. Recognized in 2020[16], spatial transcriptomics technologies, including array-based barcoding, in situ sequencing, and multiplexed hybridization, enabled comprehensive mapping of cellular organization, tissue zonation, and intercellular communication[17-22]. Building upon these frameworks, spatial proteomics has emerged as a powerful extension, leveraging technologies such as imaging mass cytometry and multiplex immunofluorescence to localize protein expression with subcellular resolution. Highlighted as a breakthrough in 2024[23], spatial proteomics enables high-resolution visualization of functional effector molecules within their native tissue context, offering unparalleled insight into the spatial organization of signaling networks, cell–cell interactions, and molecular microenvironments with both anatomical fidelity and molecular specificity[24, 25]. Spatial multi-omics enables simultaneous measurement of multiple molecular modalities, such as RNA, proteins, chromatin accessibility, and metabolites, while preserving their native positions within intact tissue[25-30]. This convergence was driven by both experimental advances and computational innovation. Barcoded microfluidic grids, imaging-based multiplex assays, and novel chemistries allowed researchers to overlay different molecular maps within the same spatial framework. By capturing molecular layers in their anatomical settings, it reveals how transcriptional circuits, regulatory programs, and metabolic states co-vary across space[31-37].

While numerous articles survey spatial transcriptomics and a growing subset profile spatial multi-omics, most emphasize experimental platforms[38-41], biological case studies[42-46], and clinical prospects[37, 47-50], with comparatively limited coverage of the computational principles that underpin rigorous analysis. In this review, we focus explicitly on modeling and algorithmic methodology for spatial multi-omics. We begin by tracing the in-depth analysis of individual spatial omics modalities, including their platforms, advantages, limitations, and data characteristics (**Figure 1**). We then discuss commonly used analytical approaches for each type of spatial omics data, including preprocessing, clustering, and data integration strategies, as well as publicly available datasets and computational resources. Finally, we highlight the transformative impact of spatial multi-omics in biomedical research and clinical medicine and discuss current challenges and future opportunities in the field.

## 2. Technological landscape of spatial multi-omics

The evolution from bulk RNA-seq to high-resolution spatial multi-omics has profoundly transformed our ability to interrogate the molecular architecture of biological systems (**Figure 2**). By preserving spatial context while capturing transcriptomic, proteomic, epigenomic, and metabolic information, these technologies offer an integrative framework for decoding the organizational principles of tissues in both health and disease. Below, in the following sections, we provide an overview of each spatial omics modality, covering their underlying technological platforms, commonly used analytical approaches, publicly available datasets, and representative applications in biological and biomedical research.

### 2.1 Spatial transcriptomics

### 2.1.1 Overview of Spatial Transcriptomics Technologies

Spatial transcriptomics has transformed genomic analysis by enabling the interrogation of gene expression within the native architectural context of tissues. Two primary technological branches dominate this domain, e.g., imaging-based approaches and sequencing-based platforms. While both share the ultimate goal of spatially resolved molecular profiling, they diverge in technical implementation, resolution, and scale of transcriptomic coverage (**Supplementary Tables 1**). This section focuses on the technologies of spatial transcriptomics, tracing its developmental trajectory from early targeted assays to high-throughput, subcellular-resolution systems capable of constructing near-complete spatial transcriptomic atlases.

Imaging-based methods fundamentally rely on the principles of in situ hybridization (ISH), offering direct visualization of RNA molecules within fixed tissue specimens. Early developments such as smFISH[51] achieved single-transcript sensitivity, laying the groundwork for spatially resolved gene expression mapping. However, these early techniques were limited by low multiplexing capabilities, typically restricted to fewer than ten genes per experiment. Subsequent innovations such as MERFISH[52] and seqFISH+ introduced combinatorial barcoding strategies coupled with sequential rounds of hybridization, significantly expanding multiplexing capacity to thousands of genes. Further progress has been driven by the pursuit of ultra-high spatial resolution and whole-transcriptome coverage. ExSeq[53] represented a key breakthrough by integrating in situ sequencing with physical tissue expansion, enabling spatial resolution down to approximately 70 nm. Similarly, STARmap PLUS[54] advanced spatial resolution to near 100 nm while preserving three-dimensional tissue integrity and allowing combined transcriptomic and proteomic analyses through error-robust encoding in complex tissue architectures. In parallel, efforts to enhance clinical applicability have led to the development of platforms such as Xenium[55]. This system integrates high spatial resolution with compatibility for FFPE-preserved human tissues and streamlined workflows through automated chemistry.

Complementary to this, sequencing-based approaches employ spatially barcoded oligonucleotide arrays to capture transcriptomes in situ. Pioneered by technologies like 10x Visium[15], these methods initially provided regional resolution but facilitated unbiased whole-transcriptome analysis. HDST[56] introduced a high-resolution framework through the use of randomly positioned ~2 μm beads embedded in a structured array. Seq-Scope[57] provides an innovative solution to previous limitations by using patterned oligonucleotide arrays created via solid-phase amplification to achieve a spatial resolution of 0.5~0.8 μm. Recent innovations, such as Slide-seqV2[58] and Stereo-seq[59], have dramatically enhanced precision to near-cellular and subcellular levels by utilizing DNA nanoball arrays or dense barcoded beads. This evolution permits genome-wide discovery without a priori gene selection, revealing novel spatial niches in developmental biology and disease pathology. However, they trade absolute resolution for potential off-target capture and lower sensitivity for low-abundance transcripts. Visium HD[60] is a recent refinement of the original 10X Visium platform, enhancing the resolution to 2-5 μm per feature. Visium HD offers relatively simple implementation and reproducibility but remains constrained by sample type restrictions and spot-based barcoding strategies, which still aggregate signals from multiple transcripts and subcellular compartments.

The chronological advancement of sequencing-based spatial transcriptomics reveals a well-defined trajectory of improvement across three fundamental dimensions: spatial resolution, tissue compatibility, and data throughput. Early platforms such as Visium and Slide-seq prioritized accessibility and experimental simplicity. Subsequent innovations, including HDST and Seq-Scope, achieved finer spatial granularity but at the cost of higher technical complexity. Recent developments, namely Visium HD and Stereo-seq, aim to balance spatial resolution with scalability and robustness. Future sequencing-based platforms are expected to integrate real-time signal acquisition, expand tissue compatibility, and enable dynamic spatiotemporal profiling.

## 2.1.2 Computational tools developed for Spatial Transcriptomics

The analysis of spatial transcriptomics data requires a suite of computational tools to process raw data, map gene expression profiles to spatial coordinates, and ensure data quality and reliability. From image preprocessing to downstream analysis, each step requires careful consideration to generate accurate and reproducible results. This section outlines key computational processes, including image processing, alignment and spatial mapping, clustering, deconvolution and cell-cell communication in spatial transcriptomics (**Table 1**, for the detailed information please see **Supplementary Tables 2**).

**2.1.2.1 Image Processing and Preprocessing**

Image processing is essential for ST methods that rely on microscopy, such as in situ hybridization-based techniques and in situ sequencing approaches. The analytical success of imaging-based ST hinges critically on robust image preprocessing and segmentation strategies. These procedures encompass illumination normalization, noise reduction, cellular and subcellular segmentation, transcript spot localization, and accurate decoding of molecular signals. A variety of software tools, ranging from general-purpose platforms such as CellProfiler[61] and ilastik[62] to specialized frameworks like SCS[63], Bento[64], and Polaris[65], have been adapted or purpose-built to address the unique computational demands of spatially resolved transcriptomic imaging.

CellProfiler serves as a versatile, open-source platform that enables non-programmers to construct modular pipelines for image analysis. It provides key functionalities for background correction and quantitative feature extraction. Ilastik complements CellProfiler by offering interactive pixel classification based on supervised machine learning, primarily through random forest classifiers. With the increasing resolution and complexity of ST data, deep learning based segmentation methods have gained prominence. The SCS algorithm integrates high-resolution transcriptomic signal maps with co-registered tissue images to delineate complex cellular geometries in datasets such as those generated by Stereo-seq. Building on this, the UCS[66] approach provides a generalized deep learning model that accommodates data heterogeneity across multiple platforms, including Xenium and MERFISH, offering scalable and transferable segmentation across modalities. For subcellular analysis, Bento offers a Python-based computational environment tailored to transcript localization and domain annotation at nanometer resolution. It enables transcript clustering, spatial proximity analysis, and subcompartmental enrichment studies. BIDCell[67] represents an evolution in biologically informed segmentation, leveraging transcriptomic features during neural network training. Polaris[65] presents a comprehensive deep learning framework tailored to high-plex RNA FISH imaging. It automates the entire processing pipeline, from transcript spot detection and classification to segmentation of cell boundaries, thereby facilitating high-throughput and reproducible analysis of image-derived spatial transcriptomics.

Barcode-array platforms with effective resolution below a cell diameter, such as Stereo-seq, do not directly observe cell outlines. In this regime, analysis proceeds through binning and aggregation. Spatial barcodes are grouped into fixed tiles, such as bin 50, to boost signal-to-noise and stabilize downstream statistics, or into adaptive bins whose size follows local molecule density to balance coverage in heterogeneous regions. On the aggregated lattice, resolution-enhancement models subdivide coarse spots into subspots using spatial priors and Bayesian smoothing, sharpening tissue boundaries and revealing subcellular gradients while preserving uncertainty. Since bins and subspots are not one-cell entities, single-cell references are then used to recover cellular composition.

As ST technologies continue to evolve toward higher resolution and greater multiplexing capacity, the requirements for preprocessing pipelines are becoming increasingly complex. A major priority will be the standardization of segmentation methods across diverse platforms and tissue types, ensuring consistency and interoperability in cross-study comparisons. Furthermore, scalability will be essential to accommodate the growing volume of high-resolution imaging data, particularly in whole-tissue or organ-wide studies. Deep learning models trained on multi-channel, biologically annotated datasets are poised to drive the development of the next generation of segmentation algorithms.

**2.1.2.2 Alignment and Spatial Mapping**

As spatial transcriptomics matures, there is a growing need to computationally align and harmonize data across tissue sections, experimental platforms, and biological conditions. Achieving spatial coherence across datasets involves reconciling both gene expression profiles and spatial coordinates. A range of algorithms has emerged to meet this challenge, encompassing optimal transport, diffeomorphic mapping, and Gaussian process–based deformation models.

One of the earliest efforts in this space is PASTE[68], a probabilistic algorithm that formulates alignment as an optimal transport problem between ST slices. By jointly considering spatial proximity and transcriptional similarity, PASTE establishes correspondences between adjacent tissue sections. While effective for linear registration, it is less capable of addressing nonlinear tissue warping introduced during histological preparation. To address this limitation, PASTE2[69] was introduced as a generalization of the original framework. It supports alignment across partially overlapping slices and enables multi-slice registration for volumetric reconstruction. GPSA[70] brings a nonparametric approach to spatial mapping by using Gaussian processes to model smooth coordinate transformations. This method excels in capturing complex tissue deformations, especially when aligning anatomically flexible regions. However, its performance can be sensitive to noise and may require careful regularization to avoid overfitting spatial variances. In contrast, STalign[71] leverages image-based registration by converting spatial transcriptomic measurements into raster images, enabling alignment via diffeomorphic transformations. This approach excels at modeling both coarse and fine tissue deformation and does not rely heavily on segmentation accuracy.

Recent advances have moved beyond pairwise registration, aiming to construct integrated three-dimensional atlases. For example, frameworks based on fused optimal transport[72], such as those employing Gromov–Wasserstein metrics, allow for the joint optimization of both structural and expression-based alignment. These models incorporate tissue morphology and gene expression simultaneously, enabling robust assembly of spatially consistent, multi-section datasets. As spatial transcriptomics evolves, the next generation of spatial alignment tools will likely adopt hybrid strategies that integrate transcriptomic, histological, and even proteomic information. Such methods must account for nonlinear tissue distortions, variability across platforms, and biological heterogeneity, while scaling to the increasingly large datasets generated by high-throughput ST platforms. The continued evolution of this computational frontier will be critical for constructing reference spatial transcriptomic atlases that support both basic research and clinical diagnostics.

### 2.1.2.3 Clustering and spatial domain analysis

Identification of discrete spatial domains and transcriptional niches within complex tissues is a central goal in spatial transcriptomics. To achieve this, researchers have developed a diverse array of computational approaches that integrate spatial coordinates with gene expression profiles, many of which are rooted in probabilistic inference, dimensionality reduction, and graph-based learning paradigms.

One notable strategy is implemented in BANKSY[73], which augments transcriptomic features by incorporating contextual information from neighboring spots, including averaged expression and orientation-based gradients. Probabilistic approaches such as BayesSpace[74], ADEPT[75], and SpatialPCA[76] embed spatial smoothness priors within low-dimensional embeddings. A distinct line of work employs graph neural networks (GNNs) to model transcriptomic neighborhoods. Methods like SpaGCN[77], STAGATE[78], GraphST[79], SpaceFlow[80], and conST define spatial graphs in which nodes represent individual tissue spots and edges encode their local connectivity. SpaGCN and STAGATE construct spatial adjacency matrices informed by either physical distance or histological morphology, enabling robust domain discovery. SpaceFlow and conST, which incorporate contrastive or multimodal learning objectives, enhance cluster coherence across heterogeneous tissues and platforms. Tools such as ADEPT[75] and SEDR[81] can efficiently process moderate-sized datasets and are widely applicable in exploratory analyses. Recent developments are pushing the boundary of ST clustering with new architectural designs. HyperGCN[82], for example, generalizes graph convolution to hypergraphs, capturing higher-order relationships among spatially distant but transcriptionally linked regions. Meanwhile, SpaGT[83] exemplifies the integration of self-attention modules with spatial priors, enabling the model to capture both local interactions and global transcriptomic structure.

Spatial clustering methodologies have matured into a diverse methodological landscape, encompassing probabilistic frameworks, graph-based and topological strategies, as well as deep learning–driven models. These approaches have become central to spatial omics analysis by enabling the delineation of cellular organization, tissue microenvironments, and intercellular relationships. Looking forward, key directions include the development of algorithms capable of scaling to organ-wide datasets, the systematic integration of multi-omics layers to capture complementary biological information, and the design of models that enhance the interpretability of latent representations while maintaining high predictive power.

**2.1.2.4 Spatially variable genes detection**

Identifying genes with spatially coordinated expression patterns, commonly referred to as spatially variable genes (SVGs), is fundamental to elucidating the structural and functional organization of complex tissues. Spatial transcriptomics platforms have motivated the development of specialized statistical frameworks capable of integrating spatial coordinates with gene expression data to uncover such patterns. These methods span probabilistic modeling, graph-based inference, and nonparametric association testing.

Statistical techniques grounded in spatial variance modeling, such as SpatialDE[84], SPARK[85], SPARK-X[86], and SOMDE[87], aim to quantify spatial dependencies in gene expression. SpatialDE[84] applies a Gaussian process to model gene-specific spatial trends. In contrast, SPARK-X accelerates inference through a non-parametric reformulation of spatial mixed models, eliminating the need for iterative likelihood estimation. SOMDE, which integrates self-organizing maps, reduces computational burden by clustering spatial coordinates prior to differential analysis. These approaches provide a principled framework for identifying genes with spatial gradients or localized enrichment.

Another class of tools draws on graph-based or image-processing analogies. MERINGUE[88] and BinSpect[89] employ spatial neighborhood matrices to evaluate expression autocorrelation via statistical tests such as Moran's I. These tools excel at identifying sharp spatial transitions or micro-domains. scGCO[90], inspired by computer vision techniques, casts gene detection as a graph-cut optimization problem over a spatial lattice. sepal introduces a graph-diffusion strategy to capture broader expression domains without relying on hard boundaries, thus enabling flexible spatial smoothing. To capture more complex or nonlinear spatial associations, nonparametric statistics have also been adopted. dCor, RV coefficient, and HSIC (Hilbert–Schmidt Independence Criterion) quantify the strength of association between expression values and spatial location vectors, independent of explicit modeling assumptions. These methods are particularly valuable in tissues with irregular architecture or where prior knowledge of spatial trends is lacking.

With the increasing resolution of spatial transcriptomics platforms, methodological innovation is moving toward hybrid and multimodal strategies. Integration with tissue morphology, spatial epigenetics, and spatial proteomics will offer new dimensions for cross-validation. Adaptive models that dynamically tune to spatial scale, tissue architecture, or gene expression sparsity are being actively pursued. Unified pipelines that merge probabilistic inference with graph-structured learning are anticipated to enhance both interpretability and biological resolution.

**2.1.2.5 Deconvolution with scRNA-seq data**

ST technologies based on sequencing often capture transcriptomes from multicellular regions, producing spot-level data that aggregate gene expression across cell types. To reconstruct the cellular makeup of each spot, computational deconvolution methods have been developed to integrate high-resolution scRNA-seq references with spatial measurements.

Early efforts in this area, such as SPOTlight[91], SpatialDWLS[92], and RCTD[93], rely on linear modeling frameworks. SPOTlight employs a topic modeling paradigm based on non-negative matrix factorization to estimate cell-type contributions across spatial locations. SpatialDWLS enhances prediction accuracy by applying dampened weighted least squares and correcting for gene-wise variability. RCTD frames deconvolution as a supervised assignment problem, aligning spatial gene profiles with known cellular transcriptomes while accounting for spot-specific noise. Beyond linear techniques, probabilistic inference models, such as stereoscope[94], DSTG[95], and Cell2location[96], offer greater flexibility by explicitly modeling count distributions and biological noise.

With the rise of deep learning, newer frameworks such as Tangram[97], gimVI[98], and SpaGE[99] take advantage of neural network architectures to learn mappings between scRNA-seq and spatial transcriptomes. Tangram trains a deep regression model to assign individual cells to spatial locations based on transcriptomic similarity. gimVI combines variational autoencoders with conditional priors to impute spatially resolved gene expression. SpaGE utilizes domain adaptation to transfer transcriptomic features across modalities, thereby enriching spatial gene prediction with neighborhood information. Optimal transport-based methods like SpaOTsc[72], novoSpaRc[100], and LIGER[101] introduce principles from geometric alignment to reconstruct tissue architecture. Other hybrid strategies, including stPlus, blend matrix completion and deep encoding with k-nearest neighbor smoothing to recover missing spatial signals, particularly in sparse or low-coverage datasets.

Future developments in the field are anticipated to focus on enhancing scalability for whole-organ imaging, refining models to integrate histological and morphological information, and enabling multimodal integration with spatial proteomics and epigenomics. Unified frameworks that can flexibly accommodate diverse spatial resolutions, tissue architectures, and molecular modalities will be critical for building high-resolution spatial atlases across developmental, physiological, and pathological states.

**2.1.2.6 Cell-cell communication using ST data**

Intercellular communication underpins a wide array of physiological processes, from tissue development to immune surveillance. With the advent of ST, it is now possible to infer signaling interactions between spatially localized cell populations. To facilitate this, numerous computational frameworks have been devised, each offering distinct strategies for integrating gene expression profiles with spatial constraints and ligand–receptor databases.

Tools initially designed for single-cell RNA-seq, such as the recent iterations of CellPhoneDB v3[102] and CellChat v2[103], have been enhanced to incorporate spatial parameters. These updates use physical proximity among cell types as an additional filter or weighting mechanism, allowing improved specificity in detecting biologically plausible L–R interactions within tissue space. Similarly, the Giotto pipeline extends L–R inference by statistically evaluating the spatial enrichment of co-expressing cell pairs, offering insight into proximity-driven signaling. Methods like COMMOT[104] utilize optimal transport theory to model signaling directionality and resource allocation across spatial domains. Several frameworks now decompose spatial influences into local versus distal effects. SVCA[105] and MISTy[106], for example, delineate the contributions of neighboring and systemic signals to gene expression variation, enabling a nuanced view of spatial influence on communication. NCEM[107] applies deep learning on graph-based representations of tissue architecture, capturing how the microenvironment shapes intercellular signaling. Recently developed approaches including NICHES[108] and DeepTalk[109] offer inference at finer granularity. NICHES calculates signaling potential at the level of individual cells or deconvolved spatial units, without relying on predefined clusters. DeepTalk, in turn, leverages attention mechanisms within graph neural networks to prioritize biologically relevant signals based on both spatial topology and molecular expression features. Platforms such as stLearn and CellNEST[110] integrate histological context into transcriptomic-based communication inference.

As ST continues to mature, communication inference tools are increasingly moving toward end-to-end integration of transcriptomic, spatial, and histological inputs. Early adaptations of scRNA-seq-based frameworks provided foundational insights, while optimal transport and graph learning–based models have elevated resolution and contextuality. The next generation of algorithms will likely integrate multiple data types, including spatial coordinates, protein localization, tissue architecture and enable dynamic modeling of signaling under varying physiological or pathological conditions. These advances hold the potential to yield high-resolution communication atlases that are both mechanistically grounded and spatially precise.

**2.1.2.7 Multi-omics integration for multi downstream tasks**

Multi-omics integration has emerged as a powerful paradigm to expand the scope and resolution of spatial biology. By combining transcriptomic, proteomic, and imaging-based measurements, researchers are now able to construct a more comprehensive representation of cellular states within tissues. Such integration not only mitigates the limitations inherent in single-modality data, but also provides complementary information that enhances the reconstruction of cellular heterogeneity, intercellular communication, and tissue organization at unprecedented precision. These advances position multi-omics fusion as a cornerstone for understanding complex biological systems and disease processes.

Within this broader context, the fusion of spatial transcriptomics with image-derived features has become a particularly active area of methodological development. Representative approaches include XFuse[111], which leverages Bayesian modeling to impute spatially resolved gene expression from histological images, and Starfysh[112], which integrates high-resolution imaging features with transcriptomic profiles to improve spatial gene expression inference and cell-type mapping. MUSE[113] introduces a deep generative framework to jointly learn from spatial transcriptomic data and histological context, thereby enhancing both data denoising and spatial resolution. Similarly, iStar[114] employs image-guided priors to improve the reconstruction of transcriptomic landscapes, while OmiCLIP[115] extends this strategy by incorporating additional modalities, aiming to achieve a more holistic view of tissue architecture.

Beyond image-based fusion, new algorithms are expanding the scope of spatial multi-omics integration. MultiMAP[116] is a dimensionality reduction and alignment framework designed to integrate multi-omics datasets, such as spatial transcriptomics, spatial proteomics, and imaging features, into a unified latent space. It operates by projecting each modality into its own lower-dimensional embedding and then aligning these embeddings through shared manifold structure, thereby preserving both intra-modal and cross-modal relationships. STvEA[117] extends this paradigm by linking spatial transcriptomics with CITE-seq–derived multi-omic profiles. Specifically, STvEA leverages antibody-derived epitope information as anchors to map cell states inferred from dissociated single-cell multi-omics back onto spatial transcriptomic coordinates. This approach not only improves cell-type resolution within spatial tissue sections but also enables the reconstruction of fine-grained tissue architectures that cannot be resolved by transcriptomic measurements alone.

The integration of spatial transcriptomics with genome-wide association studies (GWAS) enables the mapping of genetic risk variants onto tissue niches, thereby linking disease-associated loci to spatially resolved gene expression. This approach can uncover the anatomical and cellular contexts in which disease-relevant genes exert their effects. gsMap[118] (genetic score Mapping) is a computational framework developed to project GWAS trait signals onto spatial transcriptomic maps. It achieves this by scoring each spatial spot according to the enrichment of trait-associated gene expression, followed by rigorous statistical testing to localize trait-relevant tissue domains.

Despite their successes, these methods face several challenges. First, image-derived information may introduce biases, particularly when histological features do not fully correlate with gene expression. Second, scalability remains an issue for organ-scale or whole-body datasets, as many models are computationally intensive. Third, most current frameworks are limited to transcriptome–image integration, while the systematic fusion of proteomic, epigenomic, and metabolomic layers with spatial transcriptomics remains underexplored. In the coming stages, the integration of diverse molecular layers with imaging and spatial transcriptomics holds great potential. Extending current frameworks to incorporate proteomics, chromatin accessibility, or metabolomics could enable a deeper understanding of regulatory mechanisms and disease states. Moreover, advances in machine learning, particularly in multimodal representation learning, are expected to provide more interpretable and scalable solutions. Ultimately, multi-omics integration will be indispensable for building comprehensive tissue atlases and for translating spatial biology into clinical applications.

### 2.1.3 Existing spatial transcriptomics data resources

As spatial transcriptomics technologies mature, a new generation of databases has emerged to systematize, analyze, and disseminate spatially resolved molecular information (**Figure 3**). These repositories vary in data volume, scope, and emphasis, encompassing general-purpose aggregation, disease-focused curation, and multimodal integration. Understanding their respective strengths and limitations is crucial for selecting the appropriate resource for given research questions.

SpatialDB[119] and SODB[120] serve as early foundational repositories designed to centralize published spatial transcriptomics data across multiple species. Both focus on gene-level queries and basic visualization. However, SpatialDB includes manually curated datasets across five model organisms, whereas SODB adopts a broader but less annotated approach, with simpler query tools and limited integration of spatial analysis modules. STOmicsDB[121] and CROST[122] represent more expansive, analytically enriched platforms. Both support multi-species datasets, offer curated metadata, and provide web-based modules for spatial gene detection, clustering, and interaction analysis. A key distinction lies in CROST's emphasis on tumor biology and its adoption of standardized processing pipelines, which facilitates large-scale comparative studies. SORC[123] is distinct in its singular focus on cancer. Unlike general-purpose repositories, it integrates matched single-cell RNA-seq and spatial data from over 260 tissue sections, enabling joint analysis of spatial transcriptomic signatures and cellular heterogeneity in oncological contexts. SPASCER[124] and SOAR are both multi-functional platforms that support integrative analyses, but they diverge in scale and scope. SPASCER is optimized for deep integration of paired single-cell and spatial datasets, with a focus on pathway inference, regulatory network construction, and neighborhood-based spatial modeling. SOAR, in contrast, offers a broader collection of over 3,400 spatial samples across 42 tissue types and supports applications such as spatial drug target exploration, intercellular communication, and disease-specific expression atlases.

Emerging databases such as SPathDB[124] and DeepSpaceDB reflect a trend toward thematic specialization and improved user accessibility. SPathDB emphasizes spatial pathway activity mapping, facilitating functional interpretation of spatial expression patterns, whereas DeepSpaceDB focuses on providing harmonized data and user-friendly analysis for researchers without extensive programming experience. Aquila, though still under development, aims to bridge visualization and discovery by offering customizable interfaces and curated multi-organism datasets. As spatial omics becomes integral to clinical and translational applications, database platforms will also need to support reproducible workflows, API accessibility, and compliance with open data standards, thereby enabling broader adoption and collaborative innovation in spatial biology.

### 2.1.4 Hypothesis generation using spatial transcriptomics

The integration of gene expression profiles with spatial context has transformed the way biological questions are formulated, particularly in systems where tissue architecture is functionally instructive (**Figure 4**). ST enables researchers to detect transcriptional heterogeneity across intact tissues, making it uniquely suited for generating biologically grounded hypotheses in both developmental biology and oncology. By preserving the native organization of cells and tissues, ST reveals emergent patterns that are otherwise obscured in dissociated or bulk approaches.

During embryogenesis and organogenesis, cellular differentiation unfolds in tightly regulated spatial and temporal domains. ST facilitates the reconstruction of these developmental trajectories by capturing localized gene expression gradients[125], boundary-defining markers[126], and regionally enriched transcription factors[127]. Such datasets offer insights into morphogen-driven patterning processes. For instance, spatially restricted expression of canonical pathway targets (e.g., Wnt, BMP, Notch) in embryonic structures may suggest region-specific signaling activity guiding lineage allocation[128]. A working hypothesis might propose that spatially segregated progenitor populations which was demarcated by expression of transcription factors, such as Nkx2.1 or Pax6, give rise to anatomically distinct compartments under the influence of extrinsic positional cues.

Furthermore, by integrating ST with computational frameworks such as RNA

velocity, diffusion maps, or spatially aware pseudotime inference, researchers can map continuous transitions between cellular states within their physical environment. This enables formulation of hypotheses about spatial progression of fate decisions—for example, that radial glia in defined zones of the developing cortex transition into intermediate progenitors along a gradient of NeuroD expression shaped by morphogen availability.

In the context of tumor biology, ST uncovers intratumoral heterogeneity not only at the transcriptional level but also in relation to spatial proximity to immune cells, vasculature, or stromal elements. These spatial cues offer a foundation for formulating mechanistic hypotheses about tumor progression and immune evasion. For example, a spatial transcriptomic map revealing high expression of immune checkpoint molecules such as PD-L1 in regions adjacent to CD8+ T cells may suggest localized immune suppression at the tumor–immune interface. Similarly, the emergence of epithelial–mesenchymal transition (EMT) signatures at tumor margins supports the hypothesis that invasion is spatially coordinated through interaction with the surrounding stroma, possibly modulated by paracrine signals from fibroblasts or macrophages. These insights enable predictions about metastatic potential and therapy resistance, which can be subsequently validated through functional assays or spatial perturbation models.

Spatial transcriptomics also lends itself to the generation of broader classes of hypotheses. Co-localized expression of ligands and their cognate receptors in adjacent cell populations can point toward spatially constrained signaling networks. Analysis of spatial domains, defined through clustering or factor decomposition, may lead to predictions about functional microenvironments or novel cell states. When collected over developmental time points or longitudinal disease stages, ST further enables modeling of spatiotemporal regulatory dynamics.

## 2.2 Spatial Proteomics

Spatial proteomics represents a valuable approach for studying protein spatial distribution and function within tissue environments at high resolution. Unlike transcriptomic methods, which infer cellular activity from gene expression, spatial proteomics directly measures proteins, the molecular effectors involved in a wide range of cellular processes such as signaling, structural support, and metabolism. Spatial proteomics technologies can be broadly divided into two main categories: imaging-based antibody techniques and mass spectrometry based imaging techniques. These techniques allow for the spatial visualization and quantification of proteins in tissue sections by providing unique insights into the spatial distribution of proteins in their native tissue environments. By revealing the molecular mechanisms that underlie various biological processes, these technologies contribute to a better understanding of tissue architecture, disease progression, and cellular behavior. In the following sections, we will explore some of the leading imaging-based multiplexed immunofluorescence techniques, highlighting their principles, advantages, and limitations in advancing spatial proteomics research.

## 2.2.1 Imaging-Based Multiplexed Immunofluorescence

CODEX (CO-Detection by indEXing)[129], developed by Akoya Biosciences, enables simultaneous detection of up to 60 proteins using iterative cycles of antibody staining, imaging, and fluorophore stripping. Each antibody is conjugated with a unique oligonucleotide barcode, and specific fluorophore-labeled probes are applied in cycles to detect them. CODEX offers high multiplexing capability, compatibility with standard fluorescence microscopes, and subcellular resolution. However, it requires extensive antibody validation, and the iterative imaging process is susceptible to photobleaching, fluorophore crosstalk, and image registration errors.

IBEX (Iterative Bleaching Extends Multiplexity)[130] is a flexible, open-source protocol developed by the NIH that leverages cyclic immunofluorescence, photobleaching, and computational image stitching. It is compatible with off-the-shelf reagents, scalable, and cost-effective for academic labs, though it demands significant hands-on time and careful management of staining variation across cycles. Tissue-based cyclic immunofluorescence (t-CyCIF)[131] is another iterative staining and imaging strategy that allows detection of up to 60 protein targets and is optimized for formalin-fixed paraffin-embedded (FFPE) tissues. CosMx SMI, developed by NanoString, is a high-plex spatial proteomics platform that uses oligonucleotide-barcoded antibodies and optical decoding to capture spatial data at single-cell resolution. CoxMx provides simultaneous detection of hundreds of proteins and RNA, providing a detailed map of molecular tissue architecture.

## 2.2.2 Mass Spectrometry-Based Spatial Proteomics

Imaging mass cytometry (IMC)[132], developed by Fluidigm, utilizes metal isotope-conjugated antibodies and a laser ablation system coupled with time-of-flight mass spectrometry (CyTOF). IMC enables simultaneous detection of up to 40-50 markers without spectral overlap. It is compatible with FFPE and frozen sections and avoids issues with autofluorescence. However, it requires expensive instrumentation, has slower acquisition time, and provides relatively coarse spatial resolution. Multiplexed ion beam imaging (MIBI)[133] uses a primary ion beam to sputter the sample and detect secondary ions from metal-tagged antibodies. MIBI achieves higher spatial resolution than IMC and is well suited for high-resolution tissue mapping. However, MIBI has relatively low throughput, risks signal degradation with repeated scans, and depends on complex instrumentation. Matrix-assisted laser desorption ionization mass spectrometry imaging (MALDI-MSI)[134] allows for the spatial profiling of proteins, lipids and metabolites within tissue sections by utilizing a matrix to ionize samples. The technique provides detailed molecular images with spatial resolution ranging from 10 to 50 μm, making it useful for studying tissue architecture, especially for larger biomolecules and metabolites.

Micro-arrayed spatial proteomics (MASP)[135] utilizes tissue microarray technology for the analysis of proteins using mass spectrometry. The approach allows for high-throughput analysis and is suited for FFPE tissue samples, providing protein data for regions of interest (ROIs) or grid-based tissue sections. The technique can achieve high multiplexing with strong protein stability and reproducibility. SCITO-seq[136] represents a recent hybrid method that combines tissue dissociation with spatial barcoding and CyTOF analysis. It provides high-plex protein detection with spatial information, though it sacrifices precise spatial continuity due to tissue dissociation. Deep visual proteomics (DVP)[137] integrates high-resolution microscopy, AI-driven cell segmentation and laser-capture microdissection to isolate individual cells or nuclei from stained tissue sections for ultra-sensitive LC-MS/MS analysis. The approach maps quantitative proteomic profiles back onto the original image at true single-cell or subcellular resolution, enabling deep multiplexing without antibody panels, albeit with increased workflow complexity, reliance on robust segmentation models and the need for specialized MS instrumentation.

### 2.2.3 Emerging Platforms and Trends

Digital spatial profiling (DSP), offered by NanoString, employs photocleavable oligo tags attached to antibodies. Upon UV exposure, these tags are released and sequenced to quantify protein abundance in user-defined regions of interest. DSP provides spatially resolved quantification and is highly multiplexed. However, its spatial resolution is constrained by ROI-based detection, and it does not provide full-slide coverage. Proximity ligation assays (PLA), such as Duolink, allow in situ detection of protein–protein interactions with high specificity using ligation and signal amplification. These methods provide insights into protein complexes and signaling networks. Nevertheless, PLA is low throughput and difficult to multiplex, limiting its utility in large-scale spatial analyses.

### 2.2.4 computational tools and resources developed for SP

#### 2.2.4.1 Spatial proteomics image preprocessing

Typically, once spatial proteomics images have been acquired, the very first step is to identify and isolate individual cells within those images via image segmentation.

By applying automated or semi-automated segmentation algorithms, one can delineate each cell's boundaries and record its centroid coordinates. Across all antibody channels, the per-cell fluorescence intensities are then measured and collated into a cell-by-marker expression matrix, alongside each cell's spatial coordinates, which forms the foundational dataset for all downstream spatial analyses.

Also, extracting these cellular state readouts from raw images is critical in both research and clinical contexts. Traditionally, pathologists manually annotate cells, this process is accurate but labor-intensive and difficult to scale. To address this bottleneck, computer-vision-based segmentation methods have been developed. Early approaches relied on handcrafted image features (e.g., intensity thresholds, edge detection, texture segmentation), dividing images into non-overlapping regions based on grayscale, geometry, and color characteristics.

Commonly used open-source tools for image segmentation include CellProfiler and ilastik. Both platforms offer fully automated workflows—CellProfiler via modular pipelines of thresholding, morphological operations, and object-based measurement, and ilastik through interactive, machine-learning–driven pixel classification—while also allowing users to manually annotate or "paint" regions of interest to guide and refine segmentation results. These hybrid approaches make it easy to bootstrap high-quality masks on challenging images by combining human expertise with algorithmic speed. In fact, many spatial proteomics vendors now bundle their own segmentation software: for example, the CODEX workflow is supported by CODEX software, which seamlessly imports multiplexed fluorescence data, applies optimized segmentation algorithms, and exports per-cell intensity tables ready for downstream analysis.

In recent years, the advent of deep learning architectures, particularly convolutional neural networks (CNNs) and encoder–decoder models such as U-Net, has dramatically increased both the accuracy and speed of cell segmentation. These models learn directly from annotated examples, enabling them to adapt to complex cellular morphologies, variable staining patterns, and heterogeneous tissue backgrounds. As a result, even low-contrast or noisy images can yield precise cell boundary delineations. For example, DeepCell's Mesmer mode[138] exemplifies these advances in practice. Mesmer is a pretrained U-Net–based network that simultaneously leverages nuclear (e.g., DAPI) and membrane/cytoplasmic marker channels to learn joint features of cell boundaries and nuclei, allowing it to accurately segment both tightly packed and irregularly shaped cells. Because it has been trained on an extremely large and diverse collection of publicly available fluorescence images covering hundreds of cell types and tissue contexts, Mesmer generalizes well to new spatial proteomics datasets with minimal additional annotation.

**2.2.4.2 Spatial proteomics cell type annotation**

Once a cell-by-marker expression matrix has been assembled, the next critical step is cell-type annotation. Unlike spatial transcriptomics, where tens of thousands of genes are measured, spatial proteomics panels typically include only a few dozen highly specific markers chosen a priori to distinguish cell types of interest. This targeted design both reduces background noise and simplifies downstream annotation, since each channel ideally corresponds to one or more well-characterized lineage or functional markers.

CELESTA[139] is an unsupervised machine learning algorithm designed specifically for cell type annotation in multiplexed spatial proteomics data. Unlike traditional clustering- or gating-based approaches, CELESTA directly assigns individual cells to their most probable cell types by integrating marker expression profiles with spatial contextual information. This is accomplished through a principled optimization framework that incorporates prior biological knowledge without relying on manual gating or subjective thresholding. AnnoSpat[140] is a neural network–based tool developed for automated cell-type annotation and spatial pattern analysis in large-scale spatial proteomics datasets such as those from IMC and CODEX. It combines semi-supervised and supervised learning approaches to identify cell types without requiring manually labeled training data and employs point process algorithms to quantify spatial relationships between cell types. MAPS[141] is a scalable probabilistic framework designed for automated cell type assignment in spatial proteomics data. It incorporates prior knowledge of marker proteins and leverages a deep recognition neural network to perform efficient, reference-free annotation. By taking as input the expression matrix and a predefined marker list for candidate cell types, MAPS computes probabilistic cell type scores for each cell, enabling fast and interpretable classification across millions of cells in large-scale datasets. ACDC[142] is a semi-supervised annotation method originally developed for mass cytometry. It encodes prior biological knowledge through a user-defined marker–cell type table, which is transformed into landmark profiles in high-dimensional space. Using random walk–based classification, ACDC assigns cell types by propagating information from these landmarks, enabling interpretable and automated annotation.

These methods highlight the evolving landscape of spatial proteomics annotation, where integrating biological prior knowledge, spatial information and scalable machine learning is essential. CELESTA and MAPS offer unsupervised or reference-free annotation while retaining interpretability, making them well-suited for exploratory studies. AnnoSpat leverages deep learning to handle massive datasets without requiring labeled training data and ACDC provides an interpretable, marker-driven annotation, making it suitable when curated marker information is available. The choice of methods depends on the availability of marker knowledge, scale of the dataset and the balance between automation and interpretability, reflecting broader trends in spatial data analysis.

**2.2.4.3 Spatial proteomics cellular neighborhood analysis**

Spatial clustering is a fundamental analytical module in both spatial transcriptomics and spatial proteomics, aiming to summarize how different cell types are spatially organized within a tissue. After cell type identification, neighborhood analysis becomes a critical step that shifts the focus from individual cells to their local microenvironments. Rather than analyzing cells in isolation, neighborhood analysis quantifies the composition and arrangement of cells in their spatial context—uncovering recurring multicellular structures known as cellular neighborhoods.

A standard approach[143] to cellular neighborhood analysis involves manually calculating the proportions and co-occurrence patterns of annotated cell types within defined spatial windows. By calculating these metrics, researchers can segment tissues into meaningful local regions. This approach facilitates the identification of tissue microenvironments, such as immune infiltrates in inflamed areas, stromal–epithelial interfaces, or tumor–immune cell niches in cancer. These neighborhoods often reflect key biological processes, including immune responses, tissue remodeling, and disease progression.

There are also some methods designed for single-cell resolution spatial data like spatial proteomics. CytoCommunity[144] is a spatial clustering algorithm tailored for single-cell spatial proteomics. Instead of using raw expression data, it takes cell type labels of neighboring cells as input. A graph convolutional network (GCN) extracts spatial features via message passing, followed by Mincut pooling for spectral clustering into spatial communities. Its key advantage is robustness to noise, since it relies on categorical labels rather than noisy expression values. However, its performance depends on the quality of initial cell type annotations. CellLENS[145] is a deep learning framework for imaging-based spatial omics. It combines molecular profiles, spatial coordinates, and tissue images. A CNN extracts features from cell image patches, while two parallel GNNs encode image and molecular data separately. Their outputs are fused via an MLP, enabling tasks like unsupervised spatial clustering and phenotypic pattern discovery. SpatialSort[146] is a Bayesian clustering algorithm for spatial proteomics that jointly models protein expression and spatial relationships. It takes a cell-by-protein matrix and spatial graph as input, using a hidden Markov random field (HMRF) to model spatial affinity. It supports two prior modes: a marker-based reference matrix (Prior mode) and annotated anchor cells (Anchor mode). By combining spatial structure with biological priors, it improves clustering accuracy and enables label transfer across datasets. CellCharter[147] is a scalable, technology-agnostic framework for identifying and comparing cellular niches in spatial omics data. It uses VAEs for dimensionality reduction and batch correction, builds a proximity-based cell network, aggregates multi-step neighborhood features, and performs clustering via a GMM with stability-based cluster selection.

Collectively, these methods reflect a clear shift from static, rule-based neighborhood definitions toward more adaptive and data-driven spatial clustering strategies. By leveraging spatial topology, prior knowledge, and multimodal inputs, recent tools are better equipped to handle the complexity and variability of spatial proteomics data. This shift enables more robust detection of cellular communities, especially in noisy, low-marker contexts, and facilitates the discovery of subtle spatial patterns that conventional KNN-based approaches may miss. As algorithmic designs become more modular and context-aware, they also open opportunities for greater compatibility across technologies and datasets. These developments lay the groundwork for more flexible downstream analyses and motivate the need for clustering models tailored to the specific challenges of spatial proteomics.

**2.2.4.4 Multi-modal integration methods for spatial proteomics**

The previous sections focused on analysis pipelines that rely solely on spatial proteomics data. However, spatial proteomics can also benefit significantly from multi-modal integration by leveraging information from other omics modalities, particularly transcriptomics. A common strategy is to use scRNA-seq or spatial transcriptomics datasets as references to transfer biological annotations—such as cell type identities—onto spatial proteomics data. This form of knowledge transfer bridges the gap between transcript-level and protein-level measurements, enabling more comprehensive characterization of the tissue architecture. By aligning spatial proteomics with annotated reference datasets, researchers can map high-resolution molecular labels onto spatial coordinates, facilitating downstream analyses such as cellular neighborhood profiling, spatial trajectory inference, or microenvironmental heterogeneity assessment. Integrating these complementary data types enhances both the interpretability and the biological depth of spatial proteomics studies.

MaxFuse[148] is a modality-agnostic framework for cross-modal data integration, particularly suited for weakly linked features such as those between spatial proteomics and scRNA-seq. It begins by constructing fuzzy nearest-neighbor graphs within each modality and applies graph-based smoothing to enhance linked features. Initial cell matching is performed via linear assignment on smoothed features. Through iterative joint embedding, fuzzy smoothing, and reassignment, MaxFuse refines cell alignments across modalities. The final output includes matched cell pairs and shared embedding, enabling accurate label transfer and integrated spatial analysis. MDIr[149] is a semi-supervised Bayesian framework for integrating spatial proteomics data with complementary modalities, such as gene expression time-course data, to improve subcellular localization inference. Built on Gaussian mixture models, it extends prior models like TAGM and MDI to support multi-modal data integration with uncertainty quantification. By learning dataset similarity during modeling, MDIr enables flexible and principled integration across diverse data types.

scMODAL[150] is a deep generative framework for integrating unpaired single-cell multi-omics datasets with limited known feature correspondences. It employs neural networks to project cells into a shared latent space and uses generative adversarial networks (GANs) to align distributions while preserving feature topology. Cross-modal anchors are identified via mutual nearest neighbors of positively correlated features, and geometric structure is preserved using kernel-based regularization. STELLAR[151] takes an annotated spatial single-cell reference daatset and an unannotated spatial proteomics dataset as input. Using graph convolutional networks on a cell proximity graph, it can learn low-dimensional embeddings that combine molecular features and spatial context. Based on these embeddings, STELLAR assigns unannotated cells to known cell types and identifies novel cell types by leveraging an objective funcction that controls intraclass variance and uses nearest neighbors to guide the discovery of new cell types.

Together, these approaches highlight a broader trend toward more flexible and geometry-aware integration strategies that go beyond rigid feature matching (**Figure 5**). By incorporating spatial context, modeling uncertainty, and aligning across heterogeneous modalities, they improve the resolution and robustness of downstream analyses. Such developments are particularly valuable in spatial proteomics, where limited marker coverage can be compensated by information from high-dimensional references. Moreover, the ability to integrate across platforms and tissue types not only enhances annotation quality but also facilitates cross-study comparison and biological generalization. As spatial datasets grow, complexity, and diversity, the demand for scalable, interpretable, and modality-adaptive integration frameworks will only continue to rise—setting the stage for new methodological innovations and translational applications.

## 2.2.5 Overview of public database for spatial proteomics

With the rapid advancement of spatial proteomics technologies, a growing number of public databases and datasets have emerged to collect and organize large-scale spatial proteomics data (**Figure 6**), providing valuable resources to support researchers in their analyses.

Human BioMolecular Atlas Program (HuBMAP)[152] is a large-scale data consortium that hosts an extensive collection of multimodal datasets, including scRNA-seq, spatial transcriptomics, and spatial proteomics data generated using technologies such as CODEX, IMC, and MIBI. In addition to the molecular data, HuBMAP provides detailed donor clinical information and analysis results generated using standardized data processing pipelines developed by the consortium. Aquila[153] is a comprehensive database that compiles a wide range of spatial omics datasets, organized by studies across various spatial modalities, including CODEX, IMC, MIBI, and CyCIF. The platform provides access to spatial proteomics data as well as foundational spatial analyses, such as neighborhood networks, spatial autocorrelation, and spatial entropy.

SODB[154] is another large-scale database that compiles extensive spatial omics datasets, including spatial proteomics data generated using technologies such as CODEX, MIBI, IMC and CyCIF. It offers detailed cell-type annotations for all datasets and provides users with a suite of interactive visualization tools through its website. scProAtlas[155] is a specialized database dedicated to spatial proteomics, integrating datasets generated from a variety of technologies including CODEX, MIBI, IMC, and CyCIF. For most of the included datasets, scProAtlas conducts comprehensive analyses starting from raw image segmentation, followed by cell-type annotation, manual cellular neighborhood analysis, multi-modal integration, neighborhood network construction, and spatial proximity analysis. SpatialREF[156] is a manually curated spatial omics database that integrates over 9 million annotated spots from 17 tissue types across human, mouse, and drosophila. It compiles high-quality spatial data from multiple sequencing technologies, covering more than 400 domain types.

In summary, the rapid growth of spatial proteomics has been paralleled by the development of comprehensive public databases that integrate multimodal datasets, standardized analytical pipelines, and interactive visualization tools. Resources such as HuBMAP, Aquila, SODB, SpatialREF and scProAtlas not only provide access to high-quality spatial proteomic data but also support downstream analyses ranging from cell-type annotation to spatial interaction modeling. These platforms collectively serve as essential infrastructure for advancing spatial biology research, particularly in facilitating reproducibility, cross-study comparison, and functional interpretation of spatially resolved molecular data.

### 2.2.6 Generating hypothesis with spatial proteomics

With the increasing availability of spatial proteomics datasets and the emergence of computational tools and databases, researchers are now empowered to move beyond descriptive analyses toward hypothesis-driven exploration of tissue organization, cellular interactions, and disease mechanisms (**Figure 7**). Among the spatial proteomics platforms, spatial proteomics has become one of the most widely used technologies, providing high-dimensional, spatially resolved protein expression data at single-cell resolution.

A typical analysis pipeline for spatial proteomics data often begins with cell type annotation followed by neighborhood analysis, forming the foundation for downstream biological discovery. One illustrative example of hypothesis generation using spatial proteomics is the study by Qiu et al.[157], which constructed a comprehensive spatial single-cell protein atlas from 401 hepatocellular carcinoma (HCC) samples using CODEX. The authors designed a targeted antibody panel consisting of cell type–specific markers, enabling high-confidence cell annotation. Following image segmentation, protein expression matrices were manually curated to assign cell types and phenotypes based on canonical marker expression. After cell type annotation, the study proceeded to neighborhood analysis using a KNN based approach[143]. By clustering cells according to the composition of their local cellular neighborhoods, the authors identified distinct cellular neighborhood (CN) patterns that captured spatial tissue architecture. These CNs were then analyzed across the cohort to reveal inter-sample heterogeneity. Importantly, the CN profiles were linked to clinical metadata, enabling the stratification of tumors into distinct immune spatial patterns based on the local abundance of immune cells. Using a random forest model, the study further quantified the contribution of individual cell types and neighborhood structures to each immune subtype, revealing that Vimentin$^+$ macrophages and Tregs were strongly associated with immunosuppressive tumor microenvironments. To validate and refine their hypothesis, the authors integrated scRNA-seq and spatial transcriptomics data, confirming a regulatory axis between VIM$^+$ macrophages and Tregs. Ultimately, their findings led to the hypothesis that VIM$^+$ macrophages contribute to immune evasion in HCC by modulating Treg behavior within specific spatial niches—a hypothesis that emerged directly from systematic spatial proteomics analysis, cross-sample comparison, and multi-omic integration.

A compelling example is the study by Launonen et al.[158], which profiled 117 HGSC samples using t-CyCIF, before and after chemotherapy. The authors first performed cell type annotation and cellular neighborhood analysis to reveal spatial patterns of immune cells. They observed that myeloid cells formed spatial clusters—termed Myelonets—which became more prominent after treatment and were closely associated with CD8$^+$

T cell exhaustion. By integrating scRNA-seq and spatial transcriptomics, they identified the NECTIN2–TIGIT interaction as a key immunosuppressive mechanism induced by chemotherapy. Functional assays confirmed that this pathway could predict response to immune checkpoint blockade. This led to the hypothesis that chemotherapy drives spatially confined T cell exhaustion through myeloid networks, highlighting a targetable mechanism for improving immunotherapy outcomes in ovarian cancer.

In summary, these studies exemplify a systematic approach to generating biological hypotheses using spatial proteomics. The workflow typically begins with cell type annotation, often guided by predefined marker panels, followed by neighborhood analysis to define spatial microenvironments based on local cell compositions. Cross-sample comparisons of these cellular neighborhoods, especially when linked with clinical metadata, can reveal patterns of immune organization, tumor stratification, or treatment response. To move beyond spatial description, both studies integrated scRNA-seq and spatial transcriptomics to uncover molecular mechanisms underlying spatial cellular interactions. Through this integrative, multi-modal strategy, researchers were able to formulate biologically and clinically relevant hypotheses—such as the immunosuppressive role of VIM$^+$ macrophages in HCC or myeloid-driven T cell exhaustion in HGSC—ultimately leading to the identification of targetable pathways. These examples demonstrate the power of spatial proteomics not only in mapping tissue architecture but also in driving hypothesis-driven discoveries in translational research.

### 2.3.2 Spatial epigenomics

The recent emergence of spatial epigenomics technologies marks a transformative step in mapping chromatin states across intact tissues. These approaches enable the in situ interrogation of chromatin accessibility, histone modifications, and nuclear architecture, offering an essential layer of insight into how regulatory mechanisms vary across spatial domains and contribute to cellular identity and tissue function.

One of the foundational sequencing-based methods in this domain is spatially resolved ATAC-seq, which adapts transposase-accessible chromatin profiling for use on tissue sections. By coupling in situ tagmentation with spatial barcoding via microfluidic channels or bead arrays, spatial-ATAC-seq[27] allows for genome-wide assessment of open chromatin across predefined tissue coordinates. This method has been successfully applied to capture tissue-specific regulatory elements, distinguish functional zones within complex organs, and correlate chromatin landscapes with transcriptional activity, typically at near-cellular resolution. Spatial-CUT&Tag[159] provides a complementary strategy by targeting specific histone modifications or chromatin-associated proteins through antibody-guided tethering of tagmentation enzymes. Implemented within intact tissue slices, this approach generates high-resolution maps of histone marks such as H3K27me3 and H3K4me1, directly linking chromatin state to spatial localization. However, its dependence on high-affinity, well-validated antibodies and limited read complexity restricts its genome-wide applicability. sciMAP-ATAC[160] introduces an alternative strategy through spatial indexing of nuclei prior to sequencing. This method employs microbiopsy extraction to retrieve nuclei from defined regions of frozen tissue, followed by combinatorial barcoding of chromatin fragments. Although this workflow does not capture subcellular spatial resolution, it achieves robust single-nucleus chromatin profiles with coarse spatial context preserved.

More recently, imaging-based approaches have been adapted to the spatial epigenomics landscape. Techniques inspired by multiplexed RNA imaging, such as epigenomic MERFISH[161], have extended the concept of combinatorial barcoding to visualize chromatin states directly via fluorescence microscopy. These methods offer the potential to detect dozens to hundreds of epigenetic features at subcellular resolution. Although still in early developmental stages, they represent a promising route to integrate spatial localization with nuclear topology and higher-order chromatin organization. The current limitations include signal complexity, imaging throughput, and challenges in probe design and validation.

Spatial-ATAC-seq and related assays benefit from extensions of single-cell epigenomic analysis tools. ArchR is a widely adopted framework that enables spatial ATAC-seq analysis through modules for quality control, motif enrichment, gene score integration, and clustering. Similarly, Signac (a Seurat extension) allows for seamless integration of chromatin accessibility with transcriptomic features and spatial coordinates. For peak calling, tools like MACS3 remain standard, though newer pipelines often incorporate spatial smoothing kernels or Gaussian field models for spatial peak enrichment detection. For spatial-CUT&Tag data, customized versions of CUT&TagTools and downstream alignment with BWA-MEM or Bowtie2 are typical, followed by domain segmentation using ChromVAR, cisTopic, or scOpen, especially when integrating with transcriptomic data.

Future developments are expected to advance the field through improvements in spatial resolution, expansion of molecular marker panels, and tighter integration with spatial transcriptomic and proteomic datasets. Technological convergence will likely result in multi-modal pipelines capable of capturing chromatin accessibility, histone landscapes, and 3D nuclear architecture in a single experiment. Alongside these experimental advances, computational frameworks will need to evolve to support spatially resolved epigenomic integration across time points, tissue types, and species.

### 2.3.3 Spatial Metabolomics

Conventional metabolomics approaches often require tissue homogenization, leading to a loss of spatial context critical for understanding molecular heterogeneity. Spatial metabolomics overcomes this limitation by preserving in situ molecular distributions, thereby enabling functional mapping of metabolites across diverse tissue architectures. This methodology is particularly transformative in oncology, neuroscience, and developmental biology. Recognized by Nature as one of the top emerging technologies in 2022, spatial metabolomics has evolved into a multidisciplinary platform, encompassing advanced MSI modalities, multi-omics integration, and computational modeling.

Several MSI platforms have been developed, each with unique strengths and trade-offs. For instance, AFADESI-MSI enables matrix-free ionization with picogram sensitivity and spatial resolution down to 20 μm. MALDI-FTICR offers ultra-high mass accuracy and improved resolution, although performance for low metabolites can be affected by matrix interference. DESI-MSI supports analysis of FFPE tissues with spatial resolution ranging from 35 to 200 μm, though its performance is sensitive to ambient conditions. These platforms have enabled critical discoveries. AFADESI-MSI has delineated alkaloid gradients in plant roots and metabolic zoning in esophageal tumors. MALDI-FTICR has revealed cyclic IMP accumulation in fatty liver zones, while DESI-MSI combined with laser capture microdissection (LCM) has allowed correlative proteomic-metabolomic mapping.

Spatial metabolomics, primarily driven by mass spectrometry imaging, requires toolkits designed for spectral data processing, visualization, and annotation. SCiLS Lab remains one of the most comprehensive commercial solutions for MSI data, offering spatial segmentation, co-localization, and supervised classification modules tailored to histological tissue analysis. On the open-source side, MALDIquant, Cardinal, and MSItools in R provide modular environments for preprocessing and downstream spatial analysis. OpenMS, a C++ library with Python bindings, facilitates high-performance workflows across diverse MSI formats. For compound identification, METASPACE functions as a large-scale cloud annotation platform, leveraging a curated metabolite database and machine learning classifiers to validate spatially resolved spectra. MSIReader, a flexible visualization tool for both vendor-specific and open formats, supports high-throughput exploration of metabolite distributions, while Galaxy-M and Workflow4Metabolomics offer reproducible, GUI-driven pipelines for integrating MSI data with transcriptomic and proteomic layers.

Several challenges persist, including the trade-off between spatial resolution and molecular sensitivity, limited metabolite coverage at single-cell resolution, and chemical degradation of analytes in FFPE tissues. Innovations such as DISCO-based volumetric imaging and nano-DESI probes are expanding capabilities toward 3D and nanoscale resolution. AI-guided analysis using large language models holds promise for contextualizing spatial metabolite patterns within known biochemical pathways. Spatial metabolomics is rapidly transitioning from a research tool to a translational framework for precision medicine. Future advances in single-cell resolution, temporal dynamics, and biological interpretation will further enhance its utility across biomedicine, particularly in oncology and neurodegeneration.

## 3. Applications and integrated analysis of single cell and spatial multi-omics in biomedical research

Single-cell and spatial multi-omics technologies have opened new avenues for understanding tissue structure and function in both health and disease. They allow detailed mapping of tissue organization, such as the layered structure of the cortex or specific regions of the heart and spinal cord. These tools reveal cellular heterogeneity within tissues, highlighting how different cell types are arranged and interact in space. In cancer research, spatial multi-omics helps uncover how tumors grow, invade surrounding tissues, and form metastases. It also provides insight into how immune cells respond to disease, capturing spatial patterns of immune activation and inflammation. Furthermore, by mapping how cells respond to treatment within their native environment, researchers can identify drug-resistant populations and improve therapy design. Together, these advances support a more comprehensive view of biology, linking molecular information with spatial context to drive discoveries in neuroscience, oncology, and immunology (**Figure 8**).

## 3.1 Spatial multi-omics to investigate the molecular architecture of intact tissues

The advent of spatial multi-omics has profoundly expanded our capacity to investigate the molecular architecture of intact tissues. By integrating transcriptomic, epigenomic, and metabolic data with precise spatial localization, researchers are now able to chart the organization, composition, and dynamic function of diverse organs in their native contexts. These approaches are shedding light on how spatial cues govern tissue physiology and define region-specific cellular phenotypes across both developmental and adult stages.

The central nervous system, with its intricate layering and connectivity, has been a focal point for spatial multi-omic studies[162-166]. Spatial transcriptomic maps have revealed highly organized gene expression gradients within the cerebral cortex, hippocampus, and cerebellum. These spatial patterns correspond to distinct neuronal subtypes and glial populations, suggesting that gene expression is tightly coupled to anatomical function. Molecular gradients aligned with cortical lamination, for instance, have unveiled candidate regulators of neurogenesis and synaptic plasticity. The liver serves as a canonical example of spatial tissue organization, defined by radial zonation from the portal triad to the central vein. Spatial transcriptomic and proteomic profiling has elucidated how hepatocyte populations vary continuously in gene expression across this axis, enabling compartmentalized control of lipid metabolism, ammonia detoxification, and xenobiotic clearance[20, 167-169]. Spatial epigenomic data further support the presence of chromatin state differences that parallel metabolic zonation. Non-parenchymal cells, including Kupffer cells, stellate cells, and endothelial populations, also exhibit zonally restricted transcriptional programs, reinforcing the concept of niche-specific regulation within a histologically continuous tissue.

The application of spatial multi-omics in renal tissue has provided valuable insights into the segmental specialization of the nephron[19, 170-172]. Transcriptomic and chromatin accessibility analyses have demonstrated that the same epithelial lineage adopts distinct functional identities across proximal tubules, loops of Henle, and collecting ducts. These differences are reflected not only in transport and ion channel gene expression but also in spatially patterned transcription factor activity and metabolic states. Such insights are enabling a systems-level understanding of how renal function is spatially partitioned and how specific segments respond to physiological stress or early disease.

Beyond the brain, liver, and kidney, spatial multi-omic technologies have begun to chart previously unresolved features of tissues such as the heart, gastrointestinal tract, and reproductive organs. In the heart, region-specific expression of ion transporters and signaling molecules defines pacemaker zones and conduction pathways. Studies in the gut epithelium have mapped signaling cascades along the crypt–villus axis, offering insights into spatially coordinated differentiation and renewal. In the uterus, spatiotemporal transcriptomic profiling during the menstrual cycle has identified cyclical remodeling of epithelial and stromal compartments, driven by hormone-responsive regulatory programs.

## 3.2 Spatial multi-omics in development biology

The integration of spatial transcriptomics and complementary omics technologies has fundamentally redefined how developmental processes are interrogated in situ. By preserving the spatial organization of tissues while capturing molecular information at single-cell or near–single-cell resolution, spatial multi-omics provides a powerful framework for understanding morphogenesis, lineage allocation, and tissue compartmentalization during embryonic and fetal development.

In vertebrate model systems, such as zebrafish and mice, spatial transcriptomic technologies have been instrumental in capturing stage-specific tissue organization. High-resolution, whole-embryo maps have enabled the delineation of spatial domains characterized by coordinated transcriptional programs[173, 174]. For instance, dynamic gene expression patterns along the anteroposterior and dorsoventral axes have been identified in early embryos, revealing localized expression of morphogen-regulated transcription factors and signaling molecules. These spatial domains often correspond to known anatomical territories, such as the neural plate, somites, or cardiac crescent, allowing for the generation of precise hypotheses regarding region-specific developmental regulators.

As development progresses, spatial multi-omic strategies facilitate the resolution of increasingly complex organ structures. In the developing murine brain, integrated spatial transcriptomics and single-cell RNA sequencing have uncovered regionally distinct progenitor populations and maturing neuronal subtypes, aligned with structural domains such as the forebrain, midbrain, and hindbrain[175]. Similar strategies have been applied to the fetal heart, identifying region-specific gene modules involved in chamber specification, conduction system maturation, and epicardial signaling[176]. These approaches reveal both conserved and transient gene expression modules that shape tissue identity and function. In limb development, spatially mapped expression patterns of Hox genes, extracellular matrix regulators, and signaling pathway components have provided critical insights into the establishment of proximal–distal and anterior–posterior axes[177].

Recent advances have extended spatial omics to human embryonic and fetal tissues. Molecular maps of early gestational organs, including the brain, lung, and gastrointestinal tract, have revealed both conserved lineage hierarchies and human-specific features. The integration of spatial transcriptomes with computational cell-type mapping tools has enabled detailed annotation of stem cell niches, transition zones, and morphogen gradients within the intact tissue context. These data provide the foundation for studying human developmental trajectories in a spatiotemporally resolved framework, particularly in contexts where experimental manipulation is limited.

## 3.3 Spatial multi-omics in tumor microenvironment

The application of spatial multi-omics has opened new frontiers in cancer biology by enabling the direct measurement of gene expression, chromatin accessibility, protein distribution, and metabolic activity within intact tumor microenvironments. These spatially resolved data provide critical insights into the molecular heterogeneity, niche-specific signaling, and cellular interactions that define tumor progression and therapeutic resistance[42, 43, 46, 50, 178, 179].

Breast tumors exhibit marked spatial variation in immune infiltration, epithelial cell plasticity, and stromal organization[180-182]. Spatial transcriptomics has revealed that regions at the tumor–stroma interface are enriched for immune checkpoint molecules and pro-inflammatory cytokines, indicative of localized immune suppression. Adjacent stromal compartments frequently express remodeling-associated genes and myofibroblast markers, supporting a spatially organized desmoplastic response. These findings suggest that immunological and fibroblastic heterogeneity may be shaped not only by intrinsic tumor properties but also by physical proximity and microanatomical positioning.

Liver tumors present a unique model for studying spatial adaptation in a highly metabolic organ. In hepatocellular carcinoma (HCC), transcriptomic and genomic spatial profiling has uncovered zonal specialization of tumor clones, often aligned with vascular features and oxygen gradients[183]. Clonal populations with distinct mutational burdens and transcriptional signatures preferentially occupy specific territories, such as perivenous regions enriched in angiogenic signaling. Moreover, epigenomic mapping within these zones has identified chromatin accessibility changes linked to metabolic pathway activation and dedifferentiation, highlighting how spatial context informs both evolutionary trajectories and therapeutic response.

Renal tumors, particularly clear cell renal cell carcinoma (ccRCC), are characterized by extensive metabolic reprogramming and immune cell infiltration. Spatial omics approaches have revealed metabolically specialized zones within tumor tissue, including regions with elevated glycolytic flux and lactate accumulation. These zones are frequently adjacent to areas enriched in alternatively activated macrophages and exhausted T cells, suggesting localized immunometabolic crosstalk. Transcriptional and proteomic gradients across these regions implicate a role for microenvironment-driven immune modulation in shaping tumor phenotypes and progression.

Beyond tissue-specific examples, spatial omics has identified convergent principles across tumor types. In colorectal and lung cancers, spatial maps have delineated invasion fronts, necrotic cores, and perivascular niches with distinct gene expression signatures. Tumor budding regions in colorectal cancer, for example, are marked by epithelial–mesenchymal transition and WNT signaling, pointing to spatially restricted sites of metastatic competence. Meanwhile, spatial chromatin profiling in lung adenocarcinoma has revealed epigenetic heterogeneity aligned with hypoxic gradients, offering a possible explanation for spatial variation in therapy resistance.

## 4. Future prospects and challenges

The trajectory of spatial multi-omics is defined by a convergence of technological innovation, computational breakthroughs, and growing biological complexity. As the field evolves, new approaches are expanding the scope of spatial biology across dimensions, modalities, and scales, transforming it into a predictive and integrative framework for understanding health and disease (**Figure 9**).

Most current approaches treat different omics layers in isolation. The next generation of spatial studies is rapidly moving toward the simultaneous interrogation of multiple molecular layers, spanning RNA, protein, chromatin, metabolites, and lipids, within the same tissue section. Recent integrative frameworks allow for joint representation and modeling of these data types, supporting the identification of multimodal cell states and regulatory programs. As spatial barcoding chemistries and imaging strategies advance, integrated spatial assays are expected to become increasingly standardized, enabling the comprehensive dissection of tissue function and organization at single-cell resolution. Besides, large-scale foundation models trained on integrated spatial transcriptomic, proteomic, epigenomic, and imaging datasets are expected to enable the construction of virtual cells. These models aim to capture the high-dimensional molecular state of individual cells within their native spatial environment. Such virtual representations may facilitate in silico experimentation, predictive modeling of cell behavior, and the generation of testable hypotheses, ultimately serving as digital counterparts to real biological cells.

Traditional spatial omics technologies are primarily confined to two-dimensional sections, which limits their ability to capture tissue-level complexity in full anatomical context. Emerging methods are beginning to reconstruct transcriptomic and epigenomic information in three dimensions through serial sectioning combined with computational registration, enabling near-isotropic reconstruction of tissue volumes. These reconstructions will enhance our understanding of developmental trajectories, tumor invasion paths, and niche-specific signaling within organs, offering a more holistic view of biological systems.

The integration of perturbation-based experiments with spatial omics technologies is enabling unprecedented insights into how localized molecular changes reshape tissue architecture and cellular states. Computational algorithms are expected to be developed to infer causal relationships between perturbations, whether genetic, chemical, or environmental, and their spatially resolved molecular consequences. For instance, Bayesian causal inference frameworks and attention-based GNNs are being adapted to model how perturbations propagate through spatially structured cellular neighborhoods. These models can identify critical regulatory nodes that mediate local-to-global tissue responses, and are particularly valuable for understanding processes such as immune infiltration, tumor microenvironment remodeling, and regenerative responses. Recent progress has been made using graph neural ordinary differential equations, spatiotemporal variational autoencoders, and diffusion-based generative models to simulate how tissues respond to developmental cues, injury, or therapeutic interventions. These tools can predict the future states of spatial molecular maps, reconstruct trajectories of cellular movement or differentiation, and model wave-like propagation of signaling molecules or transcriptional states.

Spatial technologies are also extending into the temporal domain. By integrating time-series sampling, lineage tracing, or perturbation-based experiments, researchers are beginning to reconstruct four-dimensional atlases of development, regeneration, and disease progression. These dynamic datasets allow for the inference of spatially governed cellular transitions and the identification of critical regulatory checkpoints. When paired with computational modeling, spatial omics may be used not only to describe current tissue states, but also to simulate responses to genetic or pharmacological interventions in silico.

Spatial multi-omics is transitioning from a descriptive tool to an integrative and predictive platform for systems biology. By capturing the spatial distribution of molecular states with increasing resolution and dimensionality, and by unifying diverse modalities through AI-enabled frameworks, the field is positioned to transform our understanding of complex tissue architecture and its dysregulation in disease. These advances will underpin the next generation of diagnostics, therapeutic strategies, and regenerative interventions grounded in the spatial logic of biology. Despite this promise, key challenges remain. Accurately modeling feedback mechanisms, capturing rare events in sparsely sampled regions, and ensuring model generalizability across individuals or systems are all active areas of research. Addressing these will require interdisciplinary collaboration, combining advances in spatial technology, dynamical systems modeling, and deep learning theory.

1. Ozsolak, F. & Milos, P.M. RNA sequencing: advances, challenges and opportunities. *Nat Rev Genet* **12**, 87-98 (2011).
2. Stark, R., Grzelak, M. & Hadfield, J. RNA sequencing: the teenage years. *Nat Rev Genet* **20**, 631-656 (2019).
3. Method of the year 2013. *Nat Methods* **11**, 1 (2014).
4. Roerink, S.F. *et al.* Intra-tumour diversification in colorectal cancer at the single-cell level. *Nature* **556**, 457-462 (2018).
5. Jovic, D. *et al.* Single-cell RNA sequencing technologies and applications: A brief overview. *Clin Transl Med* **12**, e694 (2022).
6. Papalexi, E. & Satija, R. Single-cell RNA sequencing to explore immune cell heterogeneity. *Nat Rev Immunol* **18**, 35-45 (2018).
7. Kharchenko, P.V. The triumphs and limitations of computational methods for scRNA-seq. *Nat Methods* **18**, 723-732 (2021).
8. Baysoy, A., Bai, Z., Satija, R. & Fan, R. The technological landscape and applications of single-cell multi-omics. *Nature reviews. Molecular cell biology* **24**, 695-713 (2023).
9. Method of the Year 2019: Single-cell multimodal omics. *Nat Methods* **17**, 1 (2020).
10. Heumos, L. *et al.* Best practices for single-cell analysis across modalities. *Nat Rev Genet* **24**, 550-572 (2023).
11. Badia, I.M.P. *et al.* Gene regulatory network inference in the era of single-cell multi-omics. *Nat Rev Genet* **24**, 739-754 (2023).
12. Nam, A.S., Chaligne, R. & Landau, D.A. Integrating genetic and non-genetic determinants of cancer evolution by single-cell multi-omics. *Nat Rev Genet* **22**, 3-18 (2021).
13. Chen, T.Y., You, L., Hardillo, J.A.U. & Chien, M.P. Spatial Transcriptomic Technologies. *Cells* **12** (2023).
14. Anderson, A.C. *et al.* Spatial transcriptomics. *Cancer Cell* **40**, 895-900 (2022).
15. Ståhl, P.L. *et al.* Visualization and analysis of gene expression in tissue sections by spatial transcriptomics. *Science* **353**, 78-82 (2016).
16. Method of the Year 2020: spatially resolved transcriptomics. *Nature Methods* **18**, 1-1 (2021).
17. Chen, W.T. *et al.* Spatial Transcriptomics and In Situ Sequencing to Study Alzheimer's Disease. *Cell* **182**, 976-991 e919 (2020).
18. Melo Ferreira, R., Gisch, D.L. & Eadon, M.T. Spatial transcriptomics and the kidney. *Curr Opin Nephrol Hypertens* **31**, 244-250 (2022).
19. Polonsky, M. *et al.* Spatial transcriptomics defines injury specific microenvironments and cellular interactions in kidney regeneration and disease. *Nat Commun* **15**, 7010 (2024).
20. Watson, B.R. *et al.* Spatial transcriptomics of healthy and fibrotic human liver at single-cell resolution. *Nat Commun* **16**, 319 (2025).
21. Fudge, J.B. Spatial transcriptomics of the human heart. *Nat Biotechnol* **41**, 1072 (2023).
22. Hunter, M.V., Moncada, R., Weiss, J.M., Yanai, I. & White, R.M. Spatially resolved

transcriptomics reveals the architecture of the tumor-microenvironment interface. *Nat Commun* **12**, 6278 (2021).
23. Method of the Year 2024: spatial proteomics. *Nat Methods* **21**, 2195-2196 (2024).
24. Bhatia, H.S. *et al.* Spatial proteomics in three-dimensional intact specimens. *Cell* **185**, 5040-5058 e5019 (2022).
25. Lundberg, E. & Borner, G.H.H. Spatial proteomics: a powerful discovery tool for cell biology. *Nature reviews. Molecular cell biology* **20**, 285-302 (2019).
26. Alexandrov, T. Spatial metabolomics: from a niche field towards a driver of innovation. *Nat Metab* **5**, 1443-1445 (2023).
27. Deng, Y. *et al.* Spatial profiling of chromatin accessibility in mouse and human tissues. *Nature* **609**, 375-383 (2022).
28. Jin, M.Z. & Jin, W.L. Spatial epigenome-transcriptome comapping technology. *Trends Cell Biol* **33**, 449-450 (2023).
29. Zhang, D. *et al.* Spatial epigenome-transcriptome co-profiling of mammalian tissues. *Nature* **616**, 113-122 (2023).
30. Zhao, T. *et al.* Spatial genomics enables multi-modal study of clonal heterogeneity in tissues. *Nature* **601**, 85-91 (2022).
31. Hsieh, W.C. *et al.* Spatial multi-omics analyses of the tumor immune microenvironment. *J Biomed Sci* **29**, 96 (2022).
32. Wu, X. *et al.* Spatial multi-omics at subcellular resolution via high-throughput in situ pairwise sequencing. *Nat Biomed Eng* **8**, 872-889 (2024).
33. Ma, Y., Shi, W., Dong, Y., Sun, Y. & Jin, Q. Spatial Multi-Omics in Alzheimer's Disease: A Multi-Dimensional Approach to Understanding Pathology and Progression. *Curr Issues Mol Biol* **46**, 4968-4990 (2024).
34. Kiessling, P. & Kuppe, C. Spatial multi-omics: novel tools to study the complexity of cardiovascular diseases. *Genome Med* **16**, 14 (2024).
35. Nagasawa, S., Zenkoh, J., Suzuki, Y. & Suzuki, A. Spatial omics technologies for understanding molecular status associated with cancer progression. *Cancer science* **115**, 3208-3217 (2024).
36. Wu, Y., Cheng, Y., Wang, X., Fan, J. & Gao, Q. Spatial omics: Navigating to the golden era of cancer research. *Clin Transl Med* **12**, e696 (2022).
37. Liu, L. *et al.* Spatiotemporal omics for biology and medicine. *Cell* **187**, 4488-4519 (2024).
38. Hui, T., Zhou, J., Yao, M., Xie, Y. & Zeng, H. Advances in Spatial Omics Technologies. *Small Methods*, e2401171 (2025).
39. Bressan, D., Battistoni, G. & Hannon, G.J. The dawn of spatial omics. *Science* **381**, eabq4964 (2023).
40. Moses, L. & Pachter, L. Museum of spatial transcriptomics. *Nat Methods* **19**, 534-546 (2022).
41. Ren, J., Luo, S., Shi, H. & Wang, X. Spatial omics advances for in situ RNA biology. *Mol Cell* **84**, 3737-3757 (2024).
42. Jin, Y. *et al.* Advances in spatial transcriptomics and its applications in cancer research. *Molecular cancer* **23**, 129 (2024).

43. Ahn, S. & Lee, H.S. Applicability of Spatial Technology in Cancer Research. *Cancer Res Treat* **56**, 343-356 (2024).
44. Rao, A., Barkley, D., Franca, G.S. & Yanai, I. Exploring tissue architecture using spatial transcriptomics. *Nature* **596**, 211-220 (2021).
45. To, A., Yu, Z. & Sugimura, R. Recent advancement in the spatial immuno-oncology. *Semin Cell Dev Biol* **166**, 22-28 (2025).
46. Seferbekova, Z., Lomakin, A., Yates, L.R. & Gerstung, M. Spatial biology of cancer evolution. *Nat Rev Genet* **24**, 295-313 (2023).
47. Toninelli, M., Rossetti, G. & Pagani, M. Charting the tumor microenvironment with spatial profiling technologies. *Trends Cancer* **9**, 1085-1096 (2023).
48. Akhoundova, D. & Rubin, M.A. Clinical application of advanced multi-omics tumor profiling: Shaping precision oncology of the future. *Cancer Cell* **40**, 920-938 (2022).
49. Williams, C.G., Lee, H.J., Asatsuma, T., Vento-Tormo, R. & Haque, A. An introduction to spatial transcriptomics for biomedical research. *Genome Med* **14**, 68 (2022).
50. Fomitcheva-Khartchenko, A., Kashyap, A., Geiger, T. & Kaigala, G.V. Space in cancer biology: its role and implications. *Trends Cancer* **8**, 1019-1032 (2022).
51. Raj, A., van den Bogaard, P., Rifkin, S.A., van Oudenaarden, A. & Tyagi, S. Imaging individual mRNA molecules using multiple singly labeled probes. *Nat Methods* **5**, 877-879 (2008).
52. Chen, K.H., Boettiger, A.N., Moffitt, J.R., Wang, S. & Zhuang, X. RNA imaging. Spatially resolved, highly multiplexed RNA profiling in single cells. *Science* **348**, aaa6090 (2015).
53. Alon, S. *et al.* Expansion sequencing: Spatially precise in situ transcriptomics in intact biological systems. *Science* **371** (2021).
54. Zeng, H. *et al.* Integrative in situ mapping of single-cell transcriptional states and tissue histopathology in a mouse model of Alzheimer's disease. *Nat Neurosci* **26**, 430-446 (2023).
55. Marco Salas, S. *et al.* Optimizing Xenium In Situ data utility by quality assessment and best-practice analysis workflows. *Nat Methods* **22**, 813-823 (2025).
56. Vickovic, S. *et al.* High-definition spatial transcriptomics for in situ tissue profiling. *Nat Methods* **16**, 987-990 (2019).
57. Kim, Y. *et al.* Seq-Scope: repurposing Illumina sequencing flow cells for high-resolution spatial transcriptomics. *Nat Protoc* **20**, 643-689 (2025).
58. Stickels, R.R. *et al.* Highly sensitive spatial transcriptomics at near-cellular resolution with Slide-seqV2. *Nat Biotechnol* **39**, 313-319 (2021).
59. Chen, A. *et al.* Spatiotemporal transcriptomic atlas of mouse organogenesis using DNA nanoball-patterned arrays. *Cell* **185**, 1777-1792.e1721 (2022).
60. Oliveira, M.F. *et al.* High-definition spatial transcriptomic profiling of immune cell populations in colorectal cancer. *Nat Genet* **57**, 1512-1523 (2025).
61. Carpenter, A.E. *et al.* CellProfiler: image analysis software for identifying and quantifying cell phenotypes. *Genome Biol* **7**, R100 (2006).

62. Berg, S. *et al.* ilastik: interactive machine learning for (bio)image analysis. *Nat Methods* **16**, 1226-1232 (2019).
63. Chen, H., Li, D. & Bar-Joseph, Z. SCS: cell segmentation for high-resolution spatial transcriptomics. *Nat Methods* **20**, 1237-1243 (2023).
64. Mah, C.K. *et al.* Bento: a toolkit for subcellular analysis of spatial transcriptomics data. *Genome Biol* **25**, 82 (2024).
65. Chen, J. *et al.* Cell composition inference and identification of layer-specific spatial transcriptional profiles with POLARIS. *Science advances* **9**, eadd9818 (2023).
66. Chen, Y., Xu, X., Wan, X., Xiao, J. & Yang, C. UCS: A Unified Approach to Cell Segmentation for Subcellular Spatial Transcriptomics. *Small Methods* **9**, e2400975 (2025).
67. Fu, X. *et al.* BIDCell: Biologically-informed self-supervised learning for segmentation of subcellular spatial transcriptomics data. *Nat Commun* **15**, 509 (2024).
68. Zeira, R., Land, M., Strzalkowski, A. & Raphael, B.J. Alignment and integration of spatial transcriptomics data. *Nat Methods* **19**, 567-575 (2022).
69. Liu, X., Zeira, R. & Raphael, B.J. PASTE2: Partial Alignment of Multi-slice Spatially Resolved Transcriptomics Data. *bioRxiv* (2023).
70. Jones, A., Townes, F.W., Li, D. & Engelhardt, B.E. Alignment of spatial genomics data using deep Gaussian processes. *Nat Methods* **20**, 1379-1387 (2023).
71. Clifton, K. *et al.* STalign: Alignment of spatial transcriptomics data using diffeomorphic metric mapping. *Nat Commun* **14**, 8123 (2023).
72. Cang, Z. & Nie, Q. Inferring spatial and signaling relationships between cells from single cell transcriptomic data. *Nat Commun* **11**, 2084 (2020).
73. Singhal, V. *et al.* BANKSY unifies cell typing and tissue domain segmentation for scalable spatial omics data analysis. *Nat Genet* **56**, 431-441 (2024).
74. Zhao, E. *et al.* Spatial transcriptomics at subspot resolution with BayesSpace. *Nat Biotechnol* **39**, 1375-1384 (2021).
75. Hu, Y. *et al.* ADEPT: Autoencoder with differentially expressed genes and imputation for robust spatial transcriptomics clustering. *iScience* **26**, 106792 (2023).
76. Shang, L. & Zhou, X. Spatially aware dimension reduction for spatial transcriptomics. *Nat Commun* **13**, 7203 (2022).
77. Hu, J. *et al.* SpaGCN: Integrating gene expression, spatial location and histology to identify spatial domains and spatially variable genes by graph convolutional network. *Nat Methods* **18**, 1342-1351 (2021).
78. Dong, K. & Zhang, S. Deciphering spatial domains from spatially resolved transcriptomics with an adaptive graph attention auto-encoder. *Nature Communications* **13** (2022).
79. Long, Y. *et al.* Spatially informed clustering, integration, and deconvolution of spatial transcriptomics with GraphST. *Nat Commun* **14**, 1155 (2023).
80. Ren, H., Walker, B.L., Cang, Z. & Nie, Q. Identifying multicellular spatiotemporal organization of cells with SpaceFlow. *Nat Commun* **13**, 4076 (2022).

81. Xu, H. *et al.* Unsupervised spatially embedded deep representation of spatial transcriptomics. *Genome Med* **16**, 12 (2024).
82. Ma, Y., Liu, L., Zhao, Y., Hang, B. & Zhang, Y. HyperGCN: an effective deep representation learning framework for the integrative analysis of spatial transcriptomics data. *BMC Genomics* **25**, 566 (2024).
83. Bao, X., Bai, X., Liu, X., Shi, Q. & Zhang, C. Spatially informed graph transformers for spatially resolved transcriptomics. *Commun Biol* **8**, 574 (2025).
84. Svensson, V., Teichmann, S.A. & Stegle, O. SpatialDE: identification of spatially variable genes. *Nat Methods* **15**, 343-346 (2018).
85. Sun, S., Zhu, J. & Zhou, X. Statistical analysis of spatial expression patterns for spatially resolved transcriptomic studies. *Nat Methods* **17**, 193-200 (2020).
86. Zhu, J., Sun, S. & Zhou, X. SPARK-X: non-parametric modeling enables scalable and robust detection of spatial expression patterns for large spatial transcriptomic studies. *Genome Biol* **22**, 184 (2021).
87. Hao, M., Hua, K. & Zhang, X. SOMDE: a scalable method for identifying spatially variable genes with self-organizing map. *Bioinformatics* **37**, 4392-4398 (2021).
88. Miller, B.F., Bambah-Mukku, D., Dulac, C., Zhuang, X. & Fan, J. Characterizing spatial gene expression heterogeneity in spatially resolved single-cell transcriptomic data with nonuniform cellular densities. *Genome Res* **31**, 1843-1855 (2021).
89. Dries, R. *et al.* Giotto: a toolbox for integrative analysis and visualization of spatial expression data. *Genome Biol* **22**, 78 (2021).
90. Zhang, K., Feng, W. & Wang, P. Identification of spatially variable genes with graph cuts. *Nat Commun* **13**, 5488 (2022).
91. Elosua-Bayes, M., Nieto, P., Mereu, E., Gut, I. & Heyn, H. SPOTlight: seeded NMF regression to deconvolute spatial transcriptomics spots with single-cell transcriptomes. *Nucleic Acids Res* **49**, e50 (2021).
92. Dong, R. & Yuan, G.C. SpatialDWLS: accurate deconvolution of spatial transcriptomic data. *Genome Biol* **22**, 145 (2021).
93. Cable, D.M. *et al.* Robust decomposition of cell type mixtures in spatial transcriptomics. *Nat Biotechnol* **40**, 517-526 (2022).
94. Andersson, A. *et al.* Single-cell and spatial transcriptomics enables probabilistic inference of cell type topography. *Commun Biol* **3**, 565 (2020).
95. Song, Q. & Su, J. DSTG: deconvoluting spatial transcriptomics data through graph-based artificial intelligence. *Briefings in bioinformatics* **22** (2021).
96. Kleshchevnikov, V. *et al.* Cell2location maps fine-grained cell types in spatial transcriptomics. *Nat Biotechnol* **40**, 661-671 (2022).
97. Biancalani, T. *et al.* Deep learning and alignment of spatially resolved single-cell transcriptomes with Tangram. *Nat Methods* **18**, 1352-1362 (2021).
98. Romain Lopez, A.N., Maxime Langevin, Jules Samaran, Michael I. Jordan, Nir Yosef A joint model of unpaired data from scRNA-seq and spatial transcriptomics for imputing missing gene expression measurements. *ICML Workshop on Computational Biology* (2019).
99. Abdelaal, T., Mourragui, S., Mahfouz, A. & Reinders, M.J.T. SpaGE: Spatial Gene

Enhancement using scRNA-seq. *Nucleic Acids Res* **48**, e107 (2020).
100. Moriel, N. *et al.* NovoSpaRc: flexible spatial reconstruction of single-cell gene expression with optimal transport. *Nat Protoc* **16**, 4177-4200 (2021).
101. Welch, J.D. *et al.* Single-Cell Multi-omic Integration Compares and Contrasts Features of Brain Cell Identity. *Cell* **177**, 1873-1887 e1817 (2019).
102. Garcia-Alonso, L. *et al.* Mapping the temporal and spatial dynamics of the human endometrium in vivo and in vitro. *Nat Genet* **53**, 1698-1711 (2021).
103. Jin, S., Plikus, M.V. & Nie, Q. CellChat for systematic analysis of cell-cell communication from single-cell transcriptomics. *Nat Protoc* **20**, 180-219 (2025).
104. Cang, Z. *et al.* Screening cell-cell communication in spatial transcriptomics via collective optimal transport. *Nat Methods* **20**, 218-228 (2023).
105. Arnol, D., Schapiro, D., Bodenmiller, B., Saez-Rodriguez, J. & Stegle, O. Modeling Cell-Cell Interactions from Spatial Molecular Data with Spatial Variance Component Analysis. *Cell Rep* **29**, 202-211 e206 (2019).
106. Tanevski, J., Flores, R.O.R., Gabor, A., Schapiro, D. & Saez-Rodriguez, J. Explainable multiview framework for dissecting spatial relationships from highly multiplexed data. *Genome Biol* **23**, 97 (2022).
107. Fischer, D.S., Schaar, A.C. & Theis, F.J. Modeling intercellular communication in tissues using spatial graphs of cells. *Nat Biotechnol* **41**, 332-336 (2023).
108. Mason, K. *et al.* Niche-DE: niche-differential gene expression analysis in spatial transcriptomics data identifies context-dependent cell-cell interactions. *Genome Biol* **25**, 14 (2024).
109. Yang, W. *et al.* Deciphering cell-cell communication at single-cell resolution for spatial transcriptomics with subgraph-based graph attention network. *Nat Commun* **15**, 7101 (2024).
110. Zohora, F.T. *et al.* CellNEST reveals cell-cell relay networks using attention mechanisms on spatial transcriptomics. *Nat Methods* (2025).
111. Bergenstrahle, L. *et al.* Super-resolved spatial transcriptomics by deep data fusion. *Nat Biotechnol* **40**, 476-479 (2022).
112. He, S. *et al.* Starfysh integrates spatial transcriptomic and histologic data to reveal heterogeneous tumor-immune hubs. *Nat Biotechnol* (2024).
113. Bao, F. *et al.* Integrative spatial analysis of cell morphologies and transcriptional states with MUSE. *Nat Biotechnol* **40**, 1200-1209 (2022).
114. Zhang, D. *et al.* Inferring super-resolution tissue architecture by integrating spatial transcriptomics with histology. *Nat Biotechnol* **42**, 1372-1377 (2024).
115. Chen, W. *et al.* A visual-omics foundation model to bridge histopathology with spatial transcriptomics. *Nat Methods* **22**, 1568-1582 (2025).
116. Jain, M.S. *et al.* MultiMAP: dimensionality reduction and integration of multimodal data. *Genome Biology* **22** (2021).
117. Govek, K.W. *et al.* Single-cell transcriptomic analysis of mIHC images via antigen mapping. *Science advances* **7** (2021).
118. Song, L., Chen, W., Hou, J., Guo, M. & Yang, J. Spatially resolved mapping of cells associated with human complex traits. *Nature* (2025).
119. Fan, Z., Chen, R. & Chen, X. SpatialDB: a database for spatially resolved

transcriptomes. *Nucleic Acids Res* **48**, D233-D237 (2020).
120. Yuan, Z. *et al.* SODB facilitates comprehensive exploration of spatial omics data. *Nat Methods* **20**, 387-399 (2023).
121. Xu, Z. *et al.* STOmicsDB: a comprehensive database for spatial transcriptomics data sharing, analysis and visualization. *Nucleic Acids Res* **52**, D1053-D1061 (2024).
122. Wang, G. *et al.* CROST: a comprehensive repository of spatial transcriptomics. *Nucleic Acids Res* **52**, D882-D890 (2024).
123. Zhou, W. *et al.* SORC: an integrated spatial omics resource in cancer. *Nucleic Acids Res* **52**, D1429-D1437 (2024).
124. Fan, Z. *et al.* SPASCER: spatial transcriptomics annotation at single-cell resolution. *Nucleic Acids Res* **51**, D1138-D1149 (2023).
125. Klein, D. *et al.* Mapping cells through time and space with moscot. *Nature* **638**, 1065-1075 (2025).
126. Wu, Y. *et al.* Spatial multi-omics analysis of tumor-stroma boundary cell features for predicting breast cancer progression and therapy response. *Front Cell Dev Biol* **13**, 1570696 (2025).
127. Barkley, D. *et al.* Cancer cell states recur across tumor types and form specific interactions with the tumor microenvironment. *Nat Genet* **54**, 1192-1201 (2022).
128. Qu, F. *et al.* Three-dimensional molecular architecture of mouse organogenesis. *Nat Commun* **14**, 4599 (2023).
129. Goltsev, Y. *et al.* Deep Profiling of Mouse Splenic Architecture with CODEX Multiplexed Imaging. *Cell* **174**, 968-981.e915 (2018).
130. Radtke, A.J. *et al.* IBEX: an iterative immunolabeling and chemical bleaching method for high-content imaging of diverse tissues. *Nature protocols* **17**, 378-401 (2022).
131. Lin, J.R. *et al.* Highly multiplexed immunofluorescence imaging of human tissues and tumors using t-CyCIF and conventional optical microscopes. *eLife* **7** (2018).
132. Chang, Q. *et al.* Imaging Mass Cytometry. *Cytometry. Part A : the journal of the International Society for Analytical Cytology* **91**, 160-169 (2017).
133. Liu, C.C. *et al.* Reproducible, high-dimensional imaging in archival human tissue by multiplexed ion beam imaging by time-of-flight (MIBI-TOF). **102**, 762-770 (2022).
134. Aichler, M. & Walch, A. MALDI Imaging mass spectrometry: current frontiers and perspectives in pathology research and practice. *Lab Invest* **95**, 422-431 (2015).
135. Ma, M. *et al.* In-depth mapping of protein localizations in whole tissue by micro-scaffold assisted spatial proteomics (MASP). *Nature communications* **13**, 7736 (2022).
136. Hwang, B. *et al.* SCITO-seq: single-cell combinatorial indexed cytometry sequencing. *Nature methods* **18**, 903-911 (2021).
137. Mund, A. *et al.* Deep Visual Proteomics defines single-cell identity and heterogeneity. *Nature biotechnology* **40**, 1231-1240 (2022).
138. Greenwald, N.F. *et al.* Whole-cell segmentation of tissue images with human-level performance using large-scale data annotation and deep learning. **40**,

555-565 (2022).
139. Zhang, W. *et al.* Identification of cell types in multiplexed in situ images by combining protein expression and spatial information using CELESTA. **19**, 759-769 (2022).
140. Mongia, A. *et al.* AnnoSpat annotates cell types and quantifies cellular arrangements from spatial proteomics. **15**, 3744 (2024).
141. Shaban, M. *et al.* MAPS: pathologist-level cell type annotation from tissue images through machine learning. *Nature communications* **15**, 28 (2024).
142. Lee, H.C., Kosoy, R., Becker, C.E., Dudley, J.T. & Kidd, B.A. Automated cell type discovery and classification through knowledge transfer. *Bioinformatics (Oxford, England)* **33**, 1689-1695 (2017).
143. Schürch, C.M. *et al.* Coordinated cellular neighborhoods orchestrate antitumoral immunity at the colorectal cancer invasive front. **182**, 1341-1359. e1319 (2020).
144. Hu, Y. *et al.* Unsupervised and supervised discovery of tissue cellular neighborhoods from cell phenotypes. **21**, 267-278 (2024).
145. Zhu, B. *et al.* CellLENS enables cross-domain information fusion for enhanced cell population delineation in single-cell spatial omics data. *Nature immunology* **26**, 963-974 (2025).
146. Lee, E. *et al.* SpatialSort: a Bayesian model for clustering and cell population annotation of spatial proteomics data. *Bioinformatics (Oxford, England)* **39**, i131-i139 (2023).
147. Varrone, M., Tavernari, D., Santamaria-Martínez, A., Walsh, L.A. & Ciriello, G. CellCharter reveals spatial cell niches associated with tissue remodeling and cell plasticity. *Nature genetics* **56**, 74-84 (2024).
148. Chen, S. *et al.* Integration of spatial and single-cell data across modalities with weakly linked features. **42**, 1096-1106 (2024).
149. Coleman, S.D. *et al.* Semi-supervised Bayesian integration of multiple spatial proteomics datasets. 2024.2002.2008.579519 (2024).
150. Wang, G. *et al.* scMODAL: A general deep learning framework for comprehensive single-cell multi-omics data alignment with feature links. 2024.2010. 2001.616142 (2024).
151. Shimma, S. Mass Spectrometry Imaging. *Mass spectrometry (Tokyo, Japan)* **11**, A0102 (2022).
152. Jain, S. *et al.* Advances and prospects for the Human BioMolecular Atlas Program (HuBMAP). **25**, 1089-1100 (2023).
153. Zheng, Y., Chen, Y., Ding, X., Wong, K.H. & Cheung, E.J.N.A.R. Aquila: a spatial omics database and analysis platform. **51**, D827-D834 (2023).
154. Yuan, Z. *et al.* SODB facilitates comprehensive exploration of spatial omics data. **20**, 387-399 (2023).
155. Wang, T. *et al.* scProAtlas: an atlas of multiplexed single-cell spatial proteomics imaging in human tissues. gkae990 (2024).
156. Cui, T. *et al.* SpatialRef: a reference of spatial omics with known spot annotation. *Nucleic acids research* **53**, D1215-d1223 (2025).

157. Qiu, X. *et al.* Spatial single-cell protein landscape reveals vimentinhigh macrophages as immune-suppressive in the microenvironment of hepatocellular carcinoma. 1-22 (2024).
158. Launonen, I.M. *et al.* Chemotherapy induces myeloid-driven spatially confined T cell exhaustion in ovarian cancer. *Cancer cell* **42**, 2045-2063.e2010 (2024).
159. Deng, Y. *et al.* Spatial-CUT&Tag: Spatially resolved chromatin modification profiling at the cellular level. *Science* **375**, 681-686 (2022).
160. Thornton, C.A. *et al.* Spatially mapped single-cell chromatin accessibility. *Nat Commun* **12**, 1274 (2021).
161. Lu, T., Ang, C.E. & Zhuang, X. Spatially resolved epigenomic profiling of single cells in complex tissues. *Cell* **185**, 4448-4464 e4417 (2022).
162. La Manno, G. *et al.* Molecular architecture of the developing mouse brain. *Nature* **596**, 92-96 (2021).
163. Kalita-de Croft, P. *et al.* Spatial profiling technologies and applications for brain cancers. *Expert Rev Mol Diagn* **21**, 323-332 (2021).
164. Ortiz, C., Carlen, M. & Meletis, K. Spatial Transcriptomics: Molecular Maps of the Mammalian Brain. *Annu Rev Neurosci* **44**, 547-562 (2021).
165. Joglekar, A. *et al.* A spatially resolved brain region- and cell type-specific isoform atlas of the postnatal mouse brain. *Nat Commun* **12**, 463 (2021).
166. Li, Y. *et al.* Spatiotemporal transcriptome atlas reveals the regional specification of the developing human brain. *Cell* **186**, 5892-5909.e5822 (2023).
167. Ben-Moshe, S. & Itzkovitz, S. Spatial heterogeneity in the mammalian liver. *Nat Rev Gastroenterol Hepatol* **16**, 395-410 (2019).
168. Hildebrandt, F. *et al.* Spatial Transcriptomics to define transcriptional patterns of zonation and structural components in the mouse liver. *Nat Commun* **12**, 7046 (2021).
169. Wu, Y. *et al.* Spatiotemporal Immune Landscape of Colorectal Cancer Liver Metastasis at Single-Cell Level. *Cancer Discov* **12**, 134-153 (2022).
170. Wu, H. *et al.* High resolution spatial profiling of kidney injury and repair using RNA hybridization-based in situ sequencing. *Nat Commun* **15**, 1396 (2024).
171. Melo Ferreira, R. *et al.* Integration of spatial and single-cell transcriptomics localizes epithelial cell-immune cross-talk in kidney injury. *JCI Insight* **6** (2021).
172. Dixon, E.E., Wu, H., Muto, Y., Wilson, P.C. & Humphreys, B.D. Spatially Resolved Transcriptomic Analysis of Acute Kidney Injury in a Female Murine Model. *J Am Soc Nephrol* **33**, 279-289 (2022).
173. Liu, C. *et al.* Decoding the blueprints of embryo development with single-cell and spatial omics. *Semin Cell Dev Biol* **167**, 22-39 (2025).
174. Sampath Kumar, A. *et al.* Spatiotemporal transcriptomic maps of whole mouse embryos at the onset of organogenesis. *Nat Genet* **55**, 1176-1185 (2023).
175. Piwecka, M., Rajewsky, N. & Rybak-Wolf, A. Single-cell and spatial transcriptomics: deciphering brain complexity in health and disease. *Nat Rev Neurol* **19**, 346-362 (2023).
176. Farah, E.N. *et al.* Spatially organized cellular communities form the developing human heart. *Nature* **627**, 854-864 (2024).

177. Zhang, B. *et al.* A human embryonic limb cell atlas resolved in space and time. *Nature* **635**, 668-678 (2024).
178. Liang, W. *et al.* The burgeoning spatial multi-omics in human gastrointestinal cancers. *PeerJ* **12**, e17860 (2024).
179. Xu, Z. *et al.* Precision medicine in colorectal cancer: Leveraging multi-omics, spatial omics, and artificial intelligence. *Clin Chim Acta* **559**, 119686 (2024).
180. Xiong, X., Wang, X., Liu, C.C., Shao, Z.M. & Yu, K.D. Deciphering breast cancer dynamics: insights from single-cell and spatial profiling in the multi-omics era. *Biomark Res* **12**, 107 (2024).
181. Wu, S.Z. *et al.* A single-cell and spatially resolved atlas of human breast cancers. *Nat Genet* **53**, 1334-1347 (2021).
182. Pierantoni, L., Reis, R.L., Silva-Correia, J., Oliveira, J.M. & Heavey, S. Spatial-omics technologies: the new enterprise in 3D breast cancer models. *Trends Biotechnol* **41**, 1488-1500 (2023).
183. Wu, L. *et al.* An invasive zone in human liver cancer identified by Stereo-seq promotes hepatocyte-tumor cell crosstalk, local immunosuppression and tumor progression. *Cell Res* **33**, 585-603 (2023).